\documentclass[12pt]{article}
\usepackage{amssymb,amsmath,epsfig}
\allowdisplaybreaks
\begin{document}
\title{\bf Role of $\sigma R^{2}+\gamma R_{\mu\nu}T^{\mu\nu}$ Model
on Anisotropic Polytropes}

\author{M. Sharif \thanks{msharif.math@pu.edu.pk} and Arfa Waseem
\thanks{arfawaseem.pu@gmail.com}\\
Department of Mathematics, University of the Punjab,\\
Quaid-e-Azam Campus, Lahore-54590, Pakistan.}
\date{}

\maketitle

\begin{abstract}
This paper analyzes the anisotropic stellar evolution governed by
polytropic equation of state in the framework of $f(R,T,Q)$ gravity,
where $Q=R_{ab}T^{ab}$. We construct the field equations,
hydrostatic equilibrium equation and trace equation to obtain their
solutions numerically under the influence of $\sigma R^{2}+\gamma Q$
gravity model, where $\sigma$ and $\gamma$ are arbitrary constants.
We examine the dependence of various physical characteristics such
as radial/tangential pressure, energy density, anisotropic factor,
total mass and surface redshift for specific values of the model
parameters. The physical acceptability of the considered model is
discussed by verifying the validity of energy conditions, causality
condition and adiabatic index. We also study the effects arising due
to strong non-minimal matter-curvature coupling on anisotropic
polytropes. It is found that the polytropic stars are stable and
their maximum mass point lies within the required observational
Chandrasekhar limit.
\end{abstract}

\section{Introduction}

The fascinating phenomenon of gravitational collapse and its final
state have played an inspiring role for researchers in astrophysics
as well as modern cosmology. This process is responsible to produce
remnants of massive stars known as compact objects. In the
transformation of stars, a stage appears when the radiation pressure
directed outward no longer counter balances the inward directed
gravitational pull and they undergo stellar death. The final outcome
of collapse completely depends upon the original mass of the star.
Similar to other ordinary stars, compact stars are very common and
their formation lead to white dwarfs, neutron stars and black holes.
There are many evidences for the formation of white dwarfs and
neutron stars while for black holes, the first evidence is provided
by the astronomers in Andromeda galaxy and later in M104, NGC3115,
M106 and Milky Way galaxies \cite{a}.

An astronomical object comprising polytropic equation of state (EoS)
is referred as polytropic star or polytropes. The polytropic EoS
provides an excellent advancement for adiabatic systems. In the
interior of compact objects, pressure produced from the degeneracy
of electrons and neutrons exists which is used to overcome the
effects of gravity. Tooper \cite{1}, as a pioneer, inspected the
structure of relativistic objects by considering polytropic EoS for
perfect fluid distribution. Thirukkanesh and Ragel \cite{2}
investigated two cases of polytropic EoS with polytropic exponents
$2$ and $\frac{3}{2}$ to describe the properties of anisotropic
compact objects for static spherically symmetric spacetime. Herrera
et al. \cite{3} examined the conformally flat polytropic stars
satisfying anisotropic spherically symmetric fluid distribution.
They showed physical properties of these polytropes graphically for
different values of polytropic index. Ngubelanga and Maharaj
\cite{4} studied Einstein-Maxwell system of equations with
anisotropy using polytropic EoS and generated exact solutions for
different polytropic indices. Sharif and Sadiq \cite{5} extended the
work of \cite{3} for static cylindrical system and also examined the
effects of charge on anisotropic conformally flat polytropes in
spherical symmetry.

The surface gravitational redshift $(z_{s})$ is an interesting
phenomenon in which electromagnetic radiation are redshifted in
frequency in a region of strong gravitational potential. This
provides the physics of strong relation between particles in the
interior geometry of star and its EoS. For static spherical
configurations, maximum limit for the surface redshift corresponding
to isotropic as well as anisotropic fluid are obtained as
$z_{s}\leq2$ \cite{6} and $z_{s}\leq5.211$ \cite{7}, respectively.
B\"{o}hmer and Harko \cite{8} derived the upper and lower limits for
few physical parameters corresponding to anisotropic fluid with
cosmological constant. They also studied the total energy bounds and
$z_{s}$ in terms of anisotropic factor.

In the age of current cosmic accelerated expansion, alternative
theories to general relativity (GR) have played a vital role to
discover hidden mysteries of dark energy and dark matter. The $f(R)$
gravity \cite{9} is the simplest extension of GR constructed by
taking an arbitrary function $f(R)$ instead of Ricci scalar $R$ in
the Einstein-Hilbert action. Harko et al. \cite{10} proposed
$f(R,T)$ gravity, where $T$ denotes trace of the energy-momentum
tensor. The matter-curvature coupling yields a source term which may
provide interesting results. It can produce a matter dependent
deviation from geodesic motion and also describe dark energy, dark
matter interactions as well as late-time acceleration. Motivated by
this argument, Haghani et al. \cite{11} formulated a more extended
theory called $f(R,T,Q)$ gravity which involves a strong non-minimal
coupling of matter and geometry.

Odintsov and S\'{a}ez-G\'{o}mes \cite{12} analyzed various
cosmological solutions and reconstructed the corresponding
gravitational action of $f(R,T,Q)$ gravity. Sharif and Zubair
examined the validity of energy conditions \cite{13} and
thermodynamical laws \cite{14} in this theory. Ayuso et al.
\cite{15} checked the consistency and stability of this theory for
different functional forms. The physical characteristics of some
particular compact stars for different distributions of matter are
also investigated \cite{16}. Baffou et al. \cite{16a} inspected the
stability of two different models of this gravity using the de
Sitter and power-law solutions. Yousaf et al. \cite{17} explored
stability of anisotropic dissipative cylindrical system as well as
non-static anisotropic stellar models. The stability of Einstein
universe using inhomogeneous perturbations is also discussed in the
same theory \cite{17a}. Recently, we have studied non-static dust
spherical solution and analyzed the mass-radius relation as well as
redshift parameter corresponding to radial and temporal coordinates
\cite{18}.

The study of physical properties and stability of polytropic stars
is of fundamental importance in modified gravitational theories.
Henttunen et al. \cite{19} examined the stellar configurations in
$f(R)$ theory by considering different cases of polytropic EoS. They
derived the solution analytically near the center of the star and
discussed the possibility of constructing $f(R)$ theory consistent
with the solar system experiments. Orellana et al. \cite{20}
investigated the mass-radius relation as well as density profiles
using realistic polytropic EoS in $f(R)$ gravity. Henttunen and
Vilja \cite{21} observed the consistency of $f(R)$ gravity models on
polytropes for static spherical configuration. The stellar
equilibrium configuration of quark as well as polytropic stars are
studied in $f(R,T)$ gravity \cite{22}. Sharif and Siddiqa \cite{23}
analyzed the stellar structure of compact stars comprising MIT bag
model and polytropic EoS in $f(R,T)$ gravity. They also obtained the
effects of charge on static spherically symmetric polytropes for
isotropic fluid distribution \cite{24}.

From astrophysical observations, the scenario of stellar system
analyzes how stellar solution satisfies some general physical
features of compact objects. This paper is therefore devoted to
explore physical characteristics as well as stability of anisotropic
polytropes to evaluate the constraints for which the system of
stellar equations is physically realistic in $f(R,T,Q)$ gravity. The
pattern of this paper is as follows. The next section provides basic
formalism of $f(R,T,Q)$ gravity while in section \textbf{3}, we
derive equations for static spherically symmetric spacetime
corresponding to anisotropic fluid. In section \textbf{4}, we
construct a system of differential equations for two cases of
polytropic EoS and study physical features as well as stability of
compact stars graphically. We summarize the whole discussion in the
last section.

\section{Formalism of $f(R,T,Q)$ Gravity}

The basic formulation of $f(R,T,Q)$ gravity is established on the
strong association of non-minimal coupling of geometry and matter.
The action of this modified gravity in the presence of matter
Lagrangian $\mathcal{L}_{m}$ is defined as \cite{11}
\begin{equation}\label{1}
A=\frac{1}{2\kappa^{2}}\int\sqrt{-g}\left({ f(R,T,Q)+
\mathcal{L}_{m}}\right)d^4x,
\end{equation}
where $\kappa^{2}(=1)$ is the coupling constant and $g$ is the
determinant of the metric tensor $(g_{ab})$. The standard
energy-momentum tensor whose matter action depends only on $g_{ab}$
and not on its derivative is given by \cite{25}
\begin{equation}\label{2}
T_{ab}=-\frac{2}{\sqrt{-g}}\frac{\delta(\sqrt{-g}
\mathcal{L}_m)}{\delta g^{ab}}=g_{ab}
\mathcal{L}_{m}-\frac{2\partial\mathcal{L}_{m}}{\partial g^{ab}}.
\end{equation}
The corresponding field equations are
\begin{eqnarray}\nonumber
G_{ab}&=&\frac{1}{f_{R}-f_{Q}\mathcal{L}_{m}}\left[(1+f_{T}
+\frac{1}{2}Rf_{Q})T_{ab}+\left\{\frac{1}{2}(f-Rf_{R})-\mathcal{L}_{m}
f_{T}\right.\right.\\\nonumber&-&\left.\left.\frac{1}{2}\nabla_{\alpha}
\nabla_{\beta}(f_{Q}T^{\alpha\beta})\right\}g_{ab}-(g_{ab} \Box
-\nabla_{a}\nabla_{b})f_{R}+\nabla_{\alpha}\nabla_{(a}
[T_{b)}^{\alpha}f_{Q}]\right.\\\label{3}&-&\left.\frac{1}{2}
\Box(f_{Q}T_{ab})-2f_{Q}R_{\alpha(a}T_{b)}^{\alpha}
+2(f_{T}g^{\alpha\beta}+f_{Q}R^{\alpha\beta})\frac{\partial^{2}
\mathcal{L}_{m}}{\partial g^{ab}\partial g^{\alpha\beta}}\right],
\end{eqnarray}
where $f_{R}\equiv\frac{\partial f}{\partial R}$,
$f_{T}\equiv\frac{\partial f}{\partial T}$ and
$\Box\equiv\nabla^{a}\nabla_{a}$. Notice that for $Q=0$, one can
retrieve the field equations of $f(R,T)$ gravity which can further
be reduced to $f(R)$ gravity for $T=0$ and consequently, the results
of GR are obtained for $f(R)=R$. The covariant divergence of the
field equations (\ref{3}) leads to
\begin{eqnarray}\nonumber
\nabla^{a}T_{ab}&=&\frac{2}{2(1+f_{T})+Rf_{Q}}
\Big[\nabla_{a}\left(f_{Q}R^{a\mu}T_{\mu b}\right)
+\nabla_{b}(\mathcal{L}_{m}f_{T})-G_{ab}\nabla^{a}
(f_{Q}\mathcal{L}_{m})\Big.\\\label{4}&-&
\Big.\frac{1}{2}(f_{Q}R_{\alpha\beta}+f_{T}g_{\alpha\beta})
\nabla_{b} T^{\alpha\beta}-\frac{1}{2}[\nabla^{a}(Rf_{Q})
+2\nabla^{a}f_{T}]T_{ab}\Big].
\end{eqnarray}
It is mentioned here that standard conservation law does not hold in
this modified theory similar to other modified theories having
non-minimal coupling of matter and geometry \cite{10}. The trace of
the field equations (\ref{3}) is calculated in the following form
\begin{eqnarray}\nonumber
-R&=&\frac{1}{f_{R}-f_{Q}\mathcal{L}_{m}}\left[(1+\frac{1}{2}Rf_{Q})T
+2(f-Rf_{R})-\nabla_{\alpha}\nabla_{\beta}(f_{Q}T^{\alpha\beta})\right.
\\\label{5}&-&\left.3 \Box f_{R}-\frac{1}{2}\Box (T f_{Q})
-2 f_{Q} R_{\alpha\beta}T^{\alpha\beta}\right],
\end{eqnarray}

In compact objects, pressure anisotropy is an important matter
ingredient which affects their evolution. It is well-known that
stellar models are mostly rotating and anisotropic in nature. The
anisotropic factor plays a vital role in different dynamical phases
of stellar evolution \cite{25a}. The influence of anisotropy arises
when the radial component of pressure differs from the tangential or
angular component. The standard form of the energy-momentum tensor
capable of supporting anisotropic matter distribution is
\begin{equation}\label{6}
T_{ab}=(\rho + p_{t})U_{a}U_{b}+(p_{r}-p_{t})V_{a}V_{b}-
p_{t}g_{ab},
\end{equation}
where $\rho$, $p_{t}$ and $p_{r}$ represent the energy density,
tangential pressure and radial pressure of the fluid, respectively.
In comoving coordinates, $U_{a}=\sqrt{g_{00}}(1,0,0,0)$ denotes four
velocity which satisfies the condition $U_{a}U^{a}=1$ whereas
$V_{a}$ expresses the radial four-vector satisfying $V_{a}V^{a}=-1$.
For anisotropic fluid distribution, we have $\mathcal{L}_{m}=\rho$
that leads to $\frac{\partial^{2}\mathcal{L}_{m}} {\partial
g^{ab}\partial g^{\mu\nu}}=0$ \cite{11}.

In order to discuss physical characteristics of anisotropic
polytropes, we consider a particular functional form
$f(R,T,Q)=\sigma R^{m}+\gamma Q^{n}$, where $m$ and $n$ are
constants. To deal with this theory free from Ostrogradski
instabilities, we take $m\neq1$. However, this model generates
stable theory for $n = 1$ exhibiting a strong coupling between its
arguments by providing Einstein-Hilbert term involving canonical
scalar field with non-minimal variation coupling of the
energy-momentum and Ricci tensors. This functional form could
support to unveil many mysterious as well as unexplored fascinating
issues of the universe. This particular model along with $m = 2$, $n
= 1$ and constant $\gamma$ helps to study various cosmological
issues. Haghani et al. \cite{11} examined some cosmological aspects
with $n = 1$. Yousaf et al. \cite{17} discussed this model with $m
=2$, $n = 1$ and analyzed stable structure of compact objects for
anisotropic spherical distribution by taking $\gamma > 0$.
Astashenok et al. \cite{26} found that for the model $R+\sigma
R^{2}$ of $f(R)$ gravity, $\sigma>0$ yields better results to
examine the physical properties of compact stars. The case
$\sigma=0$ in $\sigma R^{m}+\gamma Q^{n}$ model gives
matter-curvature gravitational interaction only through coupling
between the Ricci and energy-momentum tensors. Such modeling could
smoothly describe dynamical implication on interesting issues of
cosmos different from $f(R,T)$ gravity.

Here, we take $m=2,~n=1,~\sigma>0$ and $\gamma>0$. Substituting
$f(R,T,Q)=\sigma R^{2}+\gamma Q$ model along with
$\mathcal{L}_{m}=\rho$ in Eq.(\ref{3}), it follows that
\begin{eqnarray}\nonumber
G_{ab}&=&\frac{1}{2\sigma R-\gamma\rho}\left[T_{ab}+\frac
{\gamma}{2}R T_{ab}-\frac{1}{2}(\sigma R^{2}-\gamma Q) g_{ab}
-\frac{\gamma}{2}\nabla_{\alpha}\nabla_{\beta}(T^{\alpha\beta})
g_{ab}\right.\\\label{7}&-&\left.2\sigma g_{ab}\Box R + 2 \sigma
\nabla_{\alpha} \nabla_{\beta} (R)+ \gamma\nabla_{\alpha}
\nabla_{(a}T_{b)}^{a}-\frac{\gamma}{2}\Box T_{ab}-2\gamma R_{\alpha
(a} T_{b)}^{\alpha}\right],
\end{eqnarray}
for which $G_{ab}$ is the usual Einstein tensor. Also, for the
considered model, the non-conservation equation of the
energy-momentum tensor yields
\begin{eqnarray}\nonumber
\nabla^{a}T_{ab}&=&\frac{2\gamma}{2+\gamma R}
\left[\nabla_{a}(R^{a\mu}T_{\mu b})-G_{ab}\nabla^{a}(
\rho)-\frac{1}{2}(R_{\alpha\beta})\nabla_{b}
T^{\alpha\beta}-\frac{1}{2}\nabla^{a}(R)T_{ab}\right].\\\label{8}
\end{eqnarray}
and the trace equation (\ref{5}) turns out to be
\begin{eqnarray}\nonumber
-R&=&\frac{1}{2\sigma R-\gamma\rho}\left[T+\frac{\gamma}{2}RT
-2\sigma R^{2}-\gamma\nabla_{\alpha}\nabla_{\beta} (T^{\alpha\beta})
-6\sigma\Box(R)-\frac{\gamma}{2}\Box (T)\right].\\\label{9}
\end{eqnarray}

\section{System of Equations}

To describe the interior geometry of a star, we take static
spherically symmetric configuration
\begin{equation}\label{10}
ds^{2}_{-}=e^{\mu(r)}dt^{2}-e^{\lambda(r)}dr^{2}
-r^{2}(d\theta^{2}+\sin^{2}\theta d\phi^{2}).
\end{equation}
For the corresponding line element, the field equations (\ref{7})
along with anisotropic matter content (\ref{6}) yield
\begin{eqnarray}\nonumber
e^{-\lambda}(\frac{\lambda'}{r}-\frac{1}{r^{2}})
+\frac{1}{r^{2}}&=&\frac{1}{2\sigma R-\gamma\rho}
\left[\rho+\frac{\gamma}{2}R\rho-\frac{\sigma}{2}R^{2}
+e^{-\lambda}\left\{\frac{\gamma}{8}\mu'^{2}\rho\right.
\right.\\\nonumber&-&\left.\left.\sigma R'
\left(\lambda'-\frac{4}{r}\right)+\gamma p_{r}
\left(\frac{\lambda'}{r}-\frac{1}{r^{2}}\right)+2\sigma R''
\right.\right.\\\nonumber&-& \left.\left.\frac{\gamma}{2}
\rho'\left(\frac{\lambda'}{2}-\frac{5\mu'}{2}
-\frac{2}{r}\right)+\frac{\gamma}{2}\left(\rho''
-p_{r}''\right)+\frac{\gamma}{r}p_{t}'\right.\right.
\\\label{11}&-&\left.\left.\frac{\gamma}{2}p_{r}'
\left(\frac{4}{r}-\frac{\lambda'}{2}\right)\right\}
\right],\\\nonumber
e^{-\lambda}\left(\frac{\mu'}{r}
+\frac{1}{r^{2}}\right)-\frac{1}{r^{2}}&=&\frac{1}{2\sigma
R-\gamma\rho}\left[p_{r} +\frac{\sigma}{2}R^{2}
+\frac{\gamma}{2}Rp_{r}+ e^{-\lambda}\left\{\frac{\gamma\rho}
{2}\right.\right.\\\nonumber&\times&\left.\left.
\left(\mu'^{2}-\mu'\lambda'\right)+\frac{\gamma}{2}
p_{r}\left(\frac{2\mu'}{r}-\mu''+\frac{6}{r^{2}}
+\lambda''\right.\right.\right.
\\\nonumber&+&\left.\left.\left.\mu'\lambda'
+\frac{4\lambda'}{r}+\frac{3\lambda'^{2}}{2}\right) +2\gamma
p_{t}\left(\frac{\lambda'}{r} -\frac{1}{r^{2}}\right)
\right.\right.\\\nonumber&+&\left.\left.\frac{\gamma}{r}p_{t}'
-\frac{\gamma}{2}\mu'\rho'+\frac{\gamma}{2}
p_{r}'\left(\frac{\mu'}{2}+\frac{2}{r}
+2\lambda'\right)\right.\right.\\\label{12}&-&\left.\left.2\sigma
R'\left(\frac{\mu'}{2}+\frac{2}{r}\right)\right\}\right],\\\nonumber
\frac{\mu''}{2}-\frac{\lambda'}{2r}+\frac{\mu'}{2r}
+\frac{\mu'^{2}}{4}-\frac{\mu'\lambda'}{4}&=&\frac{1} {2\sigma
R-\gamma\rho}\left[e^{\lambda}\left(p_{t} +\frac{\gamma}{2}
Rp_{t}+\frac{\sigma}{2}R^{2}\right) +\frac{\gamma}{4}\mu'^{2}\rho
\right.\\\nonumber&-&\left.\frac{\gamma}{2}p_{r}
\left(\frac{\mu'\lambda'}{2}-\frac{\mu'^{2}}{2}
-\mu''+\frac{2}{r^{2}}+\frac{\lambda'}{r}
-\frac{\mu'}{r}\right)\right.\\\nonumber&+&\left.\gamma
p_{t}\left(\frac{3}{r^{2}}-\frac{\mu'}{2r}\right)
+\frac{\gamma}{2}p_{r}'\left(\mu'
-\frac{\lambda'}{2}+\frac{2}{r}\right)+\frac{\gamma}{2}
p_{t}'\right.\\\nonumber&\times&\left.
\left(\frac{\mu'}{2}-\frac{\lambda'}{2}
+\frac{4}{r}\right)+2\sigma
R'\left(\frac{\lambda'}{2}-\frac{\mu'}{2}
-\frac{1}{r}\right)
\right.\\\label{13}&+&\left.\frac{\gamma}{2}
\left(p_{r}''+p_{t}''\right)-2\sigma
R''\right],
\end{eqnarray}
where prime shows derivative with respect to radial coordinate.
Moreover, from the non-conservation of the energy-momentum tensor
(\ref{8}), the hydrostatic equilibrium equation which is also a
generalized form of Tolman-Oppenheimer-Volkoff equation in
$f(R,T,Q)$ gravity is derived as
\begin{eqnarray}\nonumber
p_{r}'+\left(\rho+p_{r}\right)\frac{\mu'}{2}+\frac{2}{r}\left(p_{r}
-p_{t}\right)&=&\frac{2\gamma}{2+\gamma R} \left [p_{r} \left (\frac
{3\mu''\lambda'}{4}+\frac{\mu'\lambda''}{4}-\frac{\mu'''}{2}
-\frac{3\mu'\mu''}{4}\right.\right.\\\nonumber&+&\left.\left.
\frac{\lambda''}{r}-\frac{\mu''}{r}+\frac{\lambda'}{r^{2}}
-\frac{\mu'^{3}}{8} +\frac{\mu'\lambda'}{r}
+\frac{3\mu'^{2}\lambda'}{8}\right.\right.
\\\nonumber&-&\left.\left.\frac{\mu'\lambda'^{2}}{4}
-\frac{\lambda'^{2}}{r}\right)+\rho\left(\frac{\mu'^{2}\lambda'}{8}
-\frac{\mu'\mu''}{4}-\frac{\mu'^{2}}{2r}\right.\right.
\\\nonumber&+&\left.\left.\frac{\lambda'}{r^{2}}\right)
+p_{t}\left(\frac{2}{r^{3}}-\frac{2e^{\lambda}}{r^{3}}
-\frac{\lambda'}{r^{2}}+\frac{\mu'}{r^{2}}\right)
+R\left(\rho'\right.\right.\\\nonumber&+&
\left.\left.\frac{2}{r}(p_{r}-p_{t})\right)+\frac{p_{r}'}{2}
\left(\frac{\mu'\lambda'}{4}+\frac{\lambda'}{r}
-\frac{\mu'^{2}}{4}-\frac{\mu''}{2}\right)\right.\\\nonumber&-&
\left.\frac{\rho'}{2}\left(\frac{\mu''}{2}+\frac{\mu'}{r}
-\frac{\mu'\lambda'}{4}+\frac{\mu'^{2}}{4}\right)
+p_{t}'\left(\frac{1}{r^{2}}-\frac{e^{\lambda}}{r^{2}}
\right.\right.\\\label{14}&-&\left.\left.\frac{\lambda'}{2r}
+\frac{\mu'}{2r}\right)-\frac{R'}{2}\left(\rho-p_{r}
-2p_{t}\right)\right].
\end{eqnarray}
It is interesting to mention here that Eq.(\ref{14}) reduces to the
ideal conservation equation in GR \cite{27,28} for $\gamma=0$. For
the proposed model, the trace equation (\ref{9}) corresponding to
anisotropic distribution leads to
\begin{eqnarray}\nonumber
R''&=&\frac{e^{\lambda}}{6\sigma}\left[p_{r}+2p_{t}-\rho
+\frac{\gamma}{2}R(p_{r}+2p_{t}+\rho)\right]
-R'\left(\frac{\mu'}{2}-\frac{\lambda'}{2}+\frac{2}{r}\right)
-\frac{\gamma}{12\sigma}\\\nonumber&\times&\left(\rho''
-3p_{r}''-2p_{t}''\right)+\frac{\gamma}{12\sigma}
\rho'\left(\frac{\lambda'}{2}-\frac{2}{r}-\frac{\mu'}{2}\right)
+\frac{\gamma}{12\sigma}p_{r}'\left(\frac{5\mu'}{2}
-\frac{3\lambda'}{2}+\frac{10}{r}\right)
\\\nonumber&+&\frac{\gamma}{12\sigma}p_{t}'\left(\mu'
-\lambda'\right)+\frac{\gamma}{6\sigma}\rho\left(\frac{\mu''}{2}
+\frac{\mu'}{r}+\frac{3\mu'^{2}}{4}-\frac{\mu'\lambda'}{4}\right)
+\frac{\gamma}{6\sigma}p_{r}\left(\frac{\mu''}{2}
-\frac{\mu'\lambda'}{4}\right.\\\label{15}&-&\left.\frac{\lambda'}{r}
+\frac{2\mu'}{r}+\frac{\mu'^{2}}{4}\right)
+\frac{\gamma}{6\sigma}p_{t}\left(\frac{\lambda'}{r}
-\frac{\mu'}{r}\right).
\end{eqnarray}
In order to analyze the stellar structure of polytropic stars, we
obtain a system of differential equations with the help of
Eqs.(\ref{11})-(\ref{15}). From equations (\ref{11}) and (\ref{13}),
we obtain the following equation
\begin{eqnarray}\nonumber
\rho''&=&\rho'\left(\frac{\lambda'}{2}-\frac{5\mu'}{2}
-\frac{2}{r}\right)-2\rho\left(\frac{\mu''}{2}
+\frac{\lambda'}{2r}+\frac{\mu'}{2r}
-\frac{\mu'\lambda'}{4}+\frac{e^{\lambda}}{r^{2}}
-\frac{1}{r^{2}}+\frac{5\mu'^{2}}{8}\right)\\\nonumber&+&\frac{4\sigma
R}{\gamma}\left(\frac{\mu''}{2}+\frac{\lambda'}{2r}
+\frac{\mu'}{2r}-\frac{\mu'\lambda'}{4}+\frac{e^{\lambda}}{r^{2}}
-\frac{1}{r^{2}}+\frac{\mu'^{2}}{4}\right)
-\frac{2e^{\lambda}}{\gamma}\left(\rho+p_{t}\right)
-\frac{Re^{\lambda}}{\gamma}\\\nonumber&\times&\left(\rho
+p_{t}\right)-\frac{2\sigma R'}{\gamma}\left(\frac{2}{r}
-\mu'\right)-p_{r}\left(\mu''+\frac{\lambda'}{r}
+\frac{\mu'}{r}-\frac{\mu'\lambda'}{2}
+\frac{\mu'^{2}}{2}\right)+\frac{\mu'}{r}p_{t}
\\\label{16}&-&p_{t}''-p_{r}'\left(\mu'-\frac{2}{r}\right)
-2p_{t}'\left(\frac{3}{r}+\frac{\mu'}{4} -\frac{\lambda'}{4}\right).
\end{eqnarray}

Now we have a set of four equations, i.e., Eq.(\ref{12}) and
Eqs.(\ref{14})-(\ref{16}) with six unknowns $\rho$, $p_{r}$,
$p_{t}$, $\mu$ and $\lambda$. In order to make the system
consistent, we need EoS as well as a relation between radial and
tangential pressures which will be helpful to reduce two unknown
parameters. Heintzmann and Hillebrandt \cite{29} developed a
relation between radial and tangential pressures for a static
spherical configuration with diagonal energy-momentum tensor as
follows
\begin{eqnarray}\nonumber
p_{t}&=&p(1+\frac{\eta}{2}),\quad p_{r}=p(1-\eta),\quad
\eta<1,\\\label{17}
p_{t}&=&\frac{(1+\frac{\eta}{2})}{(1-\eta)}p_{r}=(1+\beta)p_{r}.
\end{eqnarray}
Here we are interested in the effect of anisotropic EoS on
properties of compact stars. For this purpose, we consider the above
expression to relate the radial/tangential pressure corresponding to
very simple forms of $\beta$. The constant $\beta$ is a function of
pressure only so that the anisotropy in this case arises due to the
nuclear forces on microscopic scales rather than macroscopic
deformations. We discuss the necessary conditions for the solution
of the system of equations (\ref{12}) and (\ref{14})-(\ref{16}). For
compact objects, the energy density and pressure should be finite
and regular at all points in the interior geometry. The set of
equations along with these requirements at the center of compact
objects leads to the following conditions
\begin{eqnarray}\nonumber
\mu''(0)&=&0,\quad \mu'(0)=0,\quad \mu(0)=0, \quad
\lambda'(0)=0,\quad \lambda(0)=0,\\\label{18} p_{r}'(0)&=&0,\quad
R'(0)=0,\quad R(0)=100, \quad p_{r}(0)=100,
\end{eqnarray}
where $R(0)=100$ and $p_{r}(0)=100$ are some initial values at the
center $(r=0)$ which we fix for numerical analysis. In this work, we
are taking the units of radius as $km$, mass as $M_{\odot}$ and
density (pressure) as $MeV/fm^{3}$ throughout the numerical analysis
\cite{22}.

\section{Polytropic Equation of State}

To solve the system of equations, we assume a relationship between
the energy density and pressure of the fluid which represents the
state of matter under a given set of physical conditions known as
EoS. In stars, the deficiency of degeneracy pressure of electrons
and neutrons to overcome the gravitational force leads to the
formation of white dwarfs and neutron stars, respectively. In these
compact stars, pressure against the gravitational pull has the same
origin, namely quantum pressure (Pauli principle). Polytropic stars
are self-gravitating gaseous spheres that are very helpful to
describe more realistic stellar models. To examine physical
characteristics of compact objects in $f(R,T,Q)$ gravity, we
consider polytropic EoS $p_{r} = \alpha\rho^{\nu}$ with $\alpha$
being a polytropic constant and $\nu$ as a polytropic exponent. In
the following subsections, the two cases ($\nu = 2, 5/3$) of
polytropic EoS are considered to investigate the anisotropic
spherical distribution. In general, $\nu = 5/3$ illustrates stars in
adiabatic convective equilibrium while the range $2\leq\nu\leq3$
characterizes EoS of neutron stars \cite{30}.

\subsection{Case I: $\nu=2$}

In this case, we examine the polytropic star having EoS
$p_{r}=\alpha\rho^{2}$. The system of equations (12) and (14)-(16)
for $\rho=(\frac{p_{r}}{\alpha})^{\frac{1}{2}}$ corresponding to
(\ref{17}) takes the following form
\begin{eqnarray}\nonumber
p_{r}''&=&\frac{2(\alpha p_{r})^{\frac{1}{2}}}{1+2(1+\beta)(\alpha
p_{r})^{\frac{1}{2}}}\left[p_{r}'\left\{\frac{1}{2(\alpha
p_{r})^{\frac{1}{2}}}\left(\frac{\lambda'}{2}-\frac{5\mu'}{2}
-\frac{2}{r}\right)-2(1+\beta)\left(\frac{\mu'}{4}\right.\right.\right.
\\\nonumber&-&\left.\left.\left.\frac{\lambda'}{4}+\frac{3}{r}\right)
-\mu'+\frac{2}{r}+\frac{1}{4\alpha^{\frac{1}{2}}p_{r}^{\frac{3}{2}}}\right\}
-p_{r}\left(\frac{\lambda'}{r}-\frac{\mu'\lambda'}{2}+\frac{\mu'^{2}}{2}
+\mu''+\frac{\beta\mu'}{r}\right)\right.\\\nonumber&-&\left.2\left
(\frac{p_{r}}{\alpha}\right)^{\frac{1}{2}}\left(\frac{\mu''}{2}
+\frac{\lambda'}{2r} +\frac{\mu'}{2r}-\frac{\mu'\lambda'}{4}
+\frac{e^{\lambda}}{r^{2}}-\frac{1}{r^{2}}+\frac{5\mu'^{2}}{8}\right)
-\frac{e^{\lambda}}{\gamma}\left\{\left(\frac{p_{r}}{\alpha}
\right)^{\frac{1}{2}}\right.\right.\\\nonumber&+&\left.\left.2(1
+\beta)p_{r}\right\} \left(2+R\right)+\frac{4\sigma
R}{\gamma}\left(\frac{\mu''}{2} +\frac{\lambda'}{2r}
+\frac{\mu'}{2r}-\frac{\mu'\lambda'}{4}+\frac{e^{\lambda}}{r^{2}}
-\frac{1}{r^{2}}\right.\right.\\\label{19}&+&\left.
\left.\frac{\mu'^{2}}{4}\right)-\frac{2\sigma
R'}{\gamma}\left(\frac{2}{r}-\mu'\right)\right],\\\nonumber
\lambda''&=&\lambda'\left(\frac{4\beta}{r}-\mu'
-\frac{3\lambda'}{2}-\frac{2p_{r}'}{p_{r}}+\frac{\mu'}{(\alpha
p_{r})^{\frac{1}{2}}}\right)+e^{\lambda}\left(\frac{2}{r^{2}(\alpha
p_{r})^{\frac{1}{2}}}-R-\frac{2}{\gamma}\right.\\\nonumber&-&\left.\frac{\sigma
R^{2}}{\gamma p_{r}}-\frac{4\sigma R}{\gamma
r^{2}p_{r}}\right)+\left(\frac{4\sigma R}{\gamma
rp_{r}}-\frac{2}{r(\alpha p_{r})^{\frac{1}{2}}}\right)\left(\mu'
+\frac{1}{r}\right) +\mu''-\frac{\mu'^{2}}{(\alpha
p_{r})^{\frac{1}{2}}}\\\label{20}&-&\frac{2\mu'}{r}
+p_{r}'\left(\frac{\mu'}{2\alpha^{\frac{1}{2}}p_{r}^{\frac{3}{2}}}
-\frac{2(1+\beta)}{rp_{r}}-\frac{1}{p_{r}}(\frac{2}{r}
+\frac{\mu'}{2})\right)+\frac{4\sigma R'}{\gamma p_{r}}(\frac{2}{r}
+\frac{\mu'}{2}),\\\nonumber \mu'''&=&\mu''\left(\frac{3\lambda'}{2}
-\frac{3\mu'}{2}-\frac{2}{r}-\frac{p_{r}'}{2p_{r}}
-\frac{\mu'}{2(\alpha p_{r})^{\frac{1}{2}}}
-\frac{p_{r}'}{4\alpha^{\frac{1}{2}}p_{r}^{\frac{3}{2}}}\right)
+\mu'\left\{\frac{\lambda''}{2}+\frac{2\lambda'}{r}\right.
\\\nonumber&-&\left.\frac{1}{\gamma}-\frac{\mu'^{2}}{4}
-\frac{\lambda'^{2}}{2} +\frac{3\mu'\lambda'}{4}
-R-\frac{1}{\gamma(\alpha p_{r})^{\frac{1}{2}}}
+\frac{(1+\beta)p_{r}'}{2p_{r}} +\frac{2(1
+\beta)}{r^{2}}\right.\\\nonumber&-&\left.\frac{R}{(\alpha
p_{r})^{\frac{1}{2}}}+\frac{p_{r}'}{4p_{r}}
(\lambda'-\mu')+\frac{1}{(\alpha
p_{r})^{\frac{1}{2}}}\left(\frac{\mu'\lambda'}{4}
-\frac{\mu'}{r}-\frac{\mu'^{2}}{4}\right)-\frac{p_{r}'}
{2\alpha^{\frac{1}{2}}p_{r}^{\frac{3}{2}}}\right.
\\\nonumber&\times&\left.\left(\frac{1}{r}
-\frac{\lambda'}{4}+\frac{\mu'}{4}\right)\right\}
-\frac{p_{r}'}{p_{r}}\left(\frac{2}{\gamma}+R
-\frac{\lambda'}{r}\right)+\frac{2\lambda''}{r}
+\frac{2\lambda'}{r^{2}}
+\frac{Rp_{r}'}{\alpha^{\frac{1}{2}}p_{r}^{\frac{3}{2}}}
\\\nonumber&+&R'\left(3+2\beta-\frac{1}
{(\alpha p_{r})^{\frac{1}{2}}}\right)-\frac{2\lambda'^{2}}{r}
+2(1+\beta)\left\{\frac{2}{r^{3}}-\frac{2e^{\lambda}}{r^{3}}
-\frac{\lambda'}{r^{2}}-\frac{p_{r}'}{p_{r}}\right.
\\\label{21}&\times&\left.\left(\frac{e^{\lambda}}{r^{2}}
-\frac{1}{r^{2}}+\frac{\lambda'}{2r}\right)\right\},\\\nonumber
R''&=&\frac{e^{\lambda}p_{r}}{6\sigma}\left\{3+2\beta-\frac{1}{(\alpha
p_{r})^{\frac{1}{2}}}+\frac{\gamma
R}{2}\left(3+2\beta+\frac{1}{(\alpha
p_{r})^{\frac{1}{2}}}\right)\right\} -R'\left(\frac{\mu'}{2}\right.
\\\nonumber&+&\left.\frac{2}{r}-\frac{\lambda'}{2}\right)
-\frac{\gamma}{12\sigma} p_{r}''\left(\frac{1}{2(\alpha
p_{r})^{\frac{1}{2}}}-5-2\beta\right)+\frac{\gamma}{12\sigma}
p_{r}'\left(\frac{7\mu'}{2}+\frac{10}{r}-\frac{5\lambda'}{2}
\right.\\\nonumber&+&\left.\beta(\mu'-\lambda')
+\frac{p_{r}'}{4\alpha^{\frac{1}{2}}p_{r}^{\frac{3}{2}}}+\frac{1}{2(\alpha
p_{r})^{\frac{1}{2}}}\left(\frac{\lambda'}{2}
-\frac{2}{r}-\frac{\mu'}{2}\right)\right)
+\frac{\gamma}{6\sigma}p_{r}\left(\frac{\mu''}{2}
+\frac{\mu'}{r}\right.\\\label{22}&+&\left.\frac{\mu'^{2}}{4}
-\frac{\mu'\lambda'}{4}+\frac{\beta}{r}\left(\lambda'
-\mu'\right)+\frac{1}{(\alpha
p_{r})^{\frac{1}{2}}}\left(\frac{\mu''}{2}-\frac{\mu'\lambda'}{4}
+\frac{\mu'}{r}+\frac{3\mu'^{2}}{4}\right)\right).
\end{eqnarray}

We explore physical validity of these stellar equations to analyze
basic features of compact stars. For numerical analysis, we take
constant parameters such that the energy density as well as pressure
(radial and tangential) remain positive and show maximum values at
the center of stellar object while $\beta$ is taken such that the
constant $\eta$ defined in Eq.(\ref{17}) remains less than 1. Here,
the dimension of $\sigma$ and $\gamma$ is $\frac{1}{L^{2}}$, the
dimension of $\alpha$ is $L^{2}$ whereas $\beta$ is a dimensionless
quantity. Using the initial conditions (\ref{18}), we solve
Eqs.(\ref{19})-(\ref{22}) numerically and discuss the effects of
model parameters ($\sigma$ and $\gamma$) on various physical
quantities. The behavior of radial/tangential pressure, energy
density and Ricci scalar for the considered polytropic EoS is shown
in Figure \textbf{1}.
\begin{figure}\center
\epsfig{file=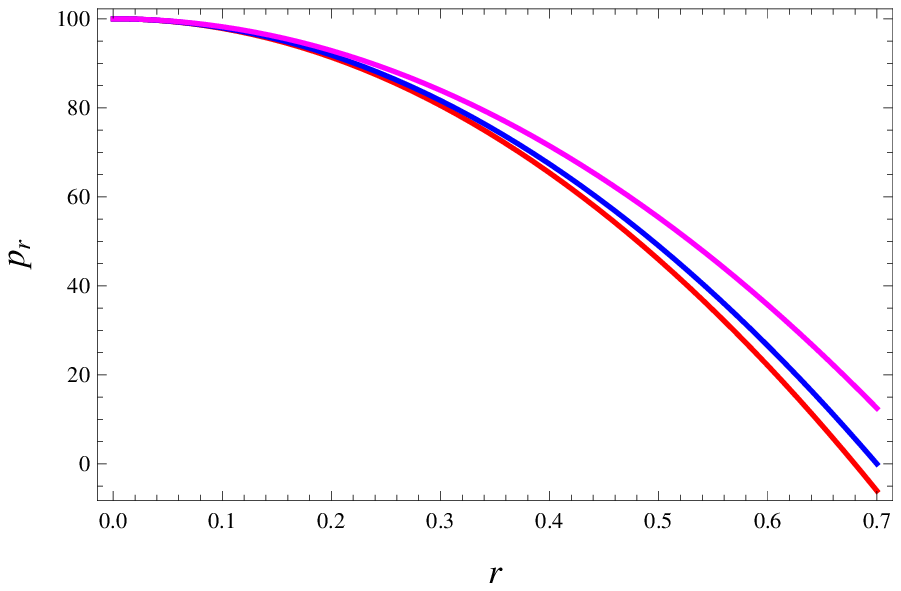,width=0.4\linewidth}\epsfig{file=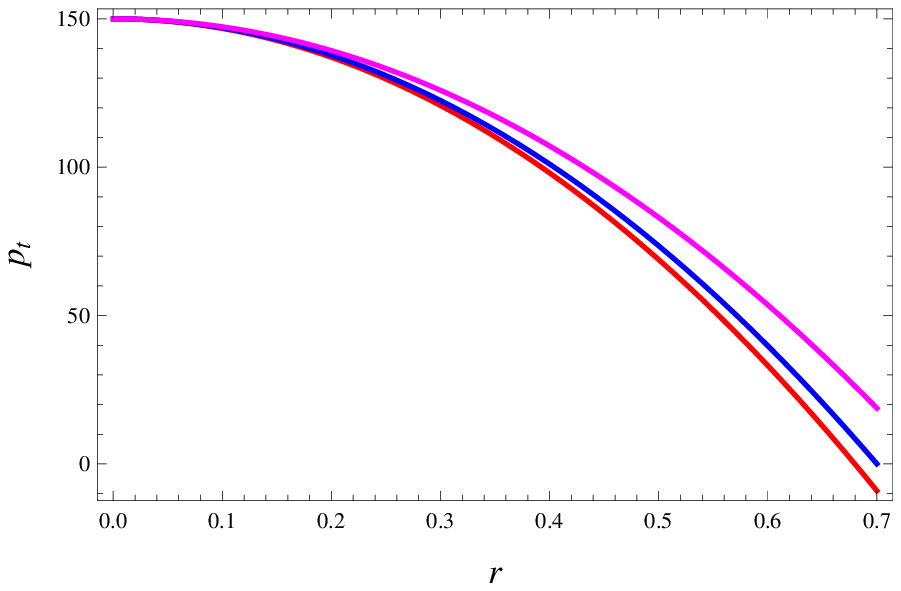,width=0.4\linewidth}
\epsfig{file=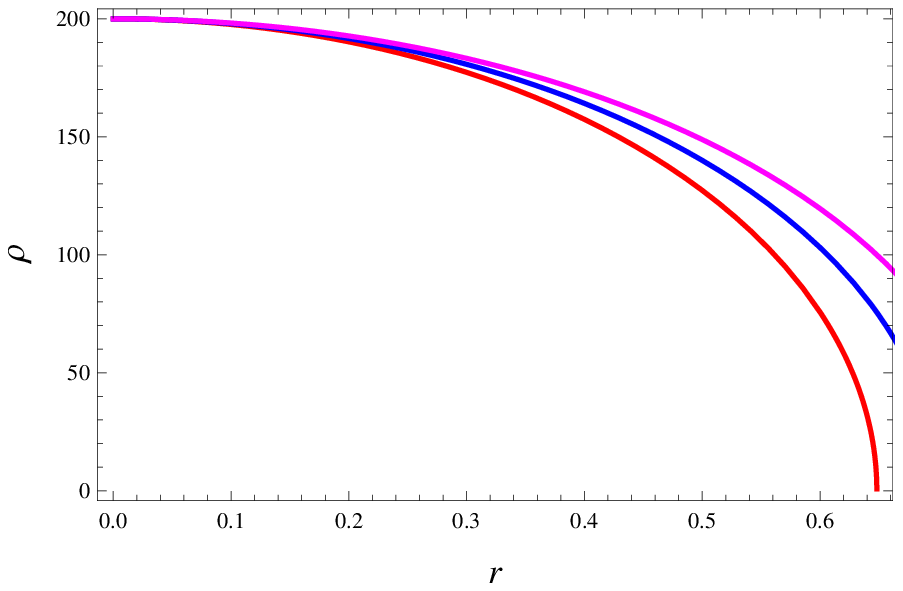,width=0.4\linewidth}\epsfig{file=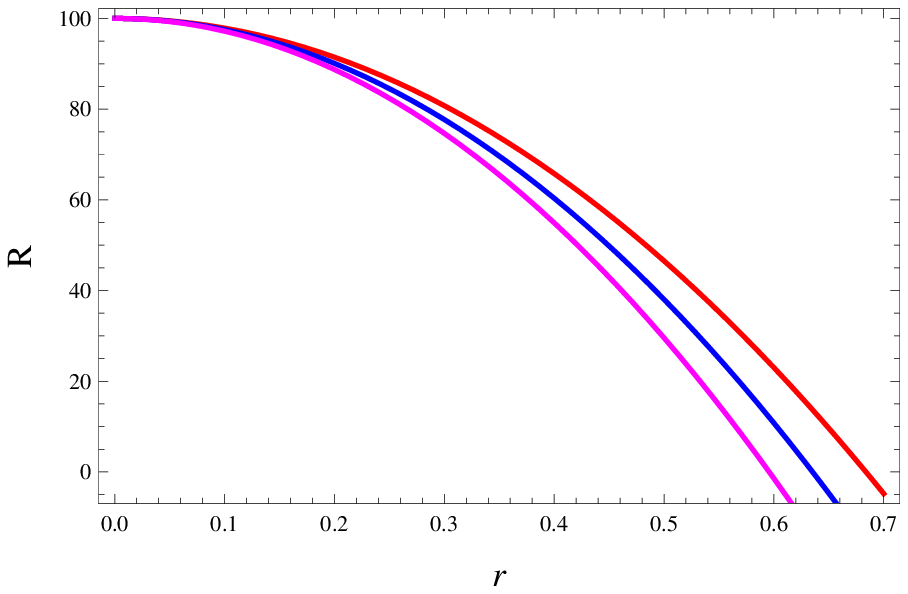,width=0.4\linewidth}
\caption{Variation of $p_{r}$, $p_{t}$, $\rho$ and $R$ versus $r$
for $p_{r}=\alpha\rho^{2}$, $\beta=0.5$, $\alpha=0.0025$,
$\sigma=2$, $\gamma=30$ (red), $\gamma=35$ (blue) and $\gamma=40$
(magenta).}
\end{figure}

Figure \textbf{1} represents that pressure (radial and tangential)
and energy density are maximum at the center and their values
decrease with the increase in radial coordinate and increase with
the larger values of model parameter $\gamma$. The radial pressure
is zero at $r=0.7 km$ providing that this is the radius of a
polytropic star in this case. The plot of Ricci scalar shows
decreasing behavior with the increase in the radius of star while
its value also decreases for larger values of the coupling constant.

To investigate the presence of exotic/normal matter distribution in
the interior of polytropic star, the energy conditions for
anisotropic fluid are defined by \cite{31}
\begin{itemize}
\item Null energy condition: \quad$\rho+ p_{r}\geq 0$,\quad$\rho
+ p_{t}\geq 0$,
\item Strong energy condition: \quad$\rho+ p_{r}\geq 0$,\quad$\rho
+ p_{t}\geq0$,\quad$\rho+ p_{r}+ 2p_{t}\geq 0$,
\item Dominant energy condition: \quad$\rho\geq| p_{r}|$,\quad$\rho\geq|
p_{t}|$,
\item Weak energy condition:\quad$\rho\geq 0$,\quad$\rho+ p_{r}\geq 0$,
\quad$\rho+ p_{t}\geq0$.
\end{itemize}
The plots of all energy conditions are given in Figure \textbf{2}
which indicate that our system of equations corresponding to this
case of polytropic EoS is feasible with all the energy conditions
for different parametric values of $\gamma$. This analysis also
ensures physical viability of our proposed functional form of this
gravity.
\begin{figure}\center
\epsfig{file=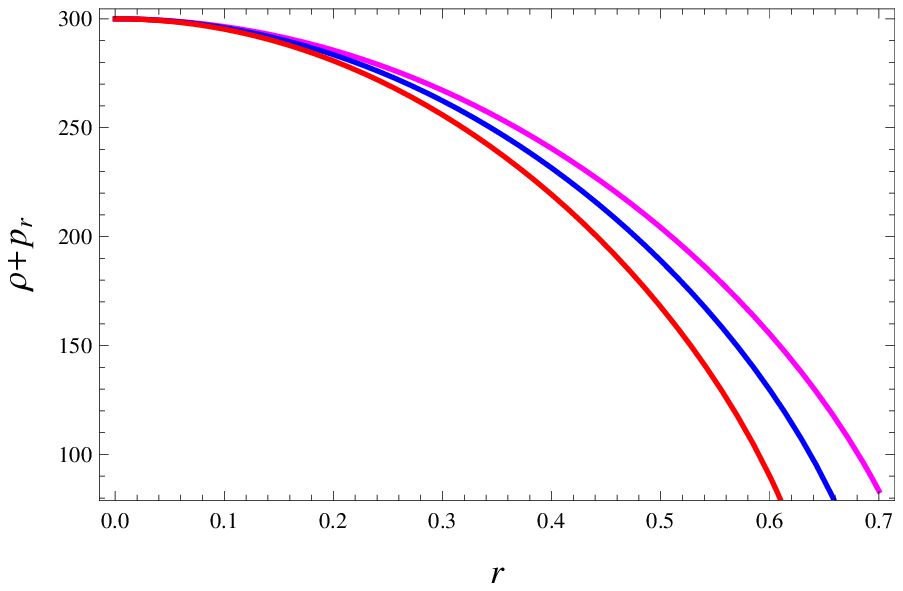,width=0.5\linewidth}\epsfig{file=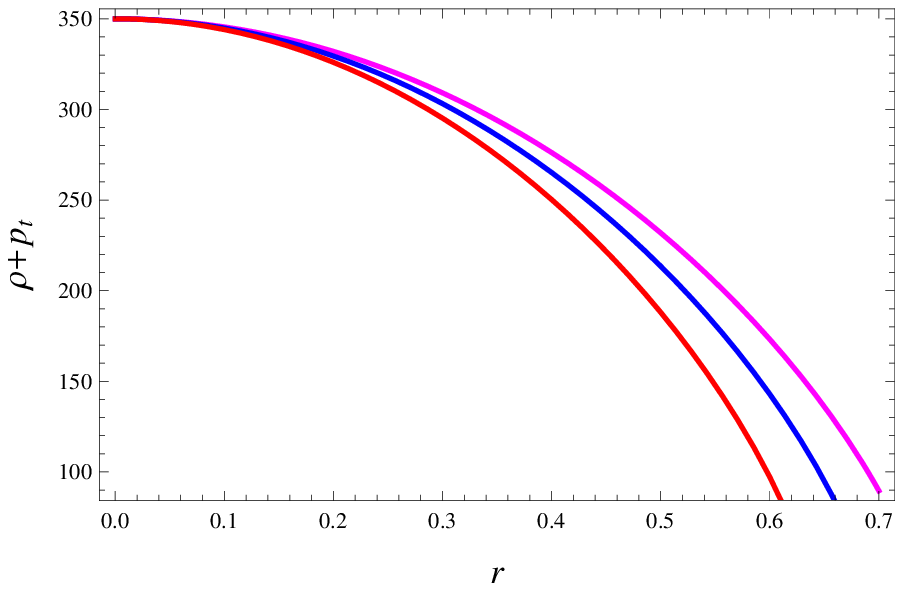,width=0.5\linewidth}
\epsfig{file=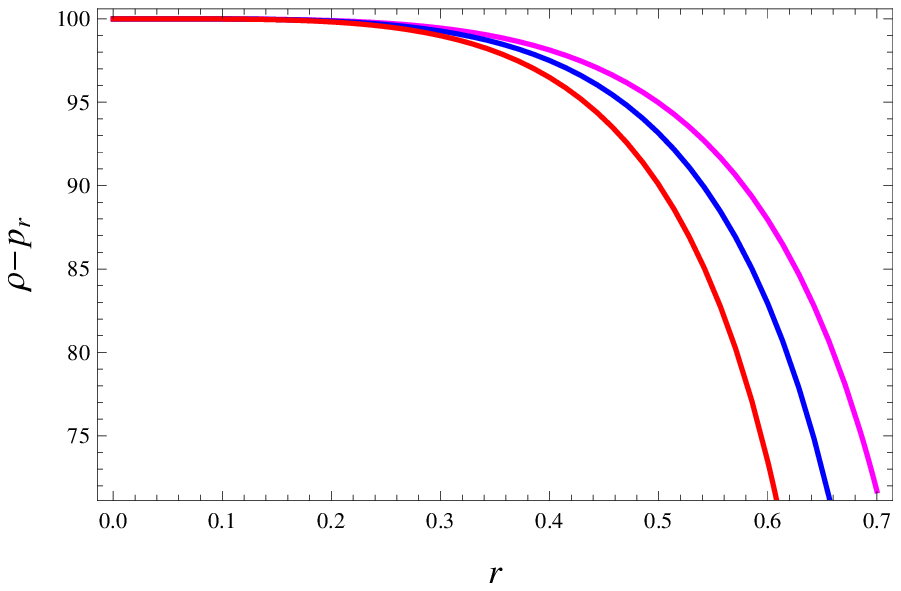,width=0.5\linewidth}\epsfig{file=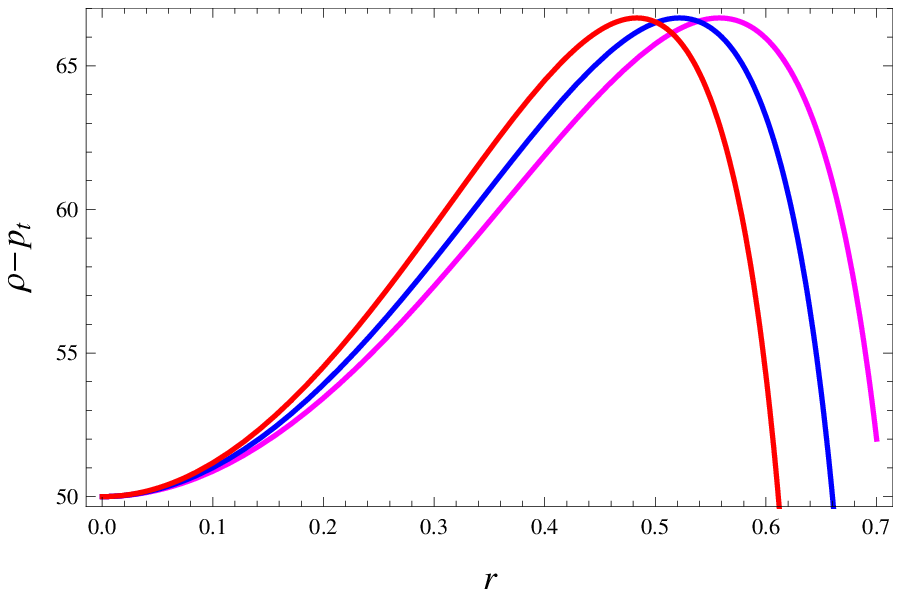,width=0.5\linewidth}
\epsfig{file=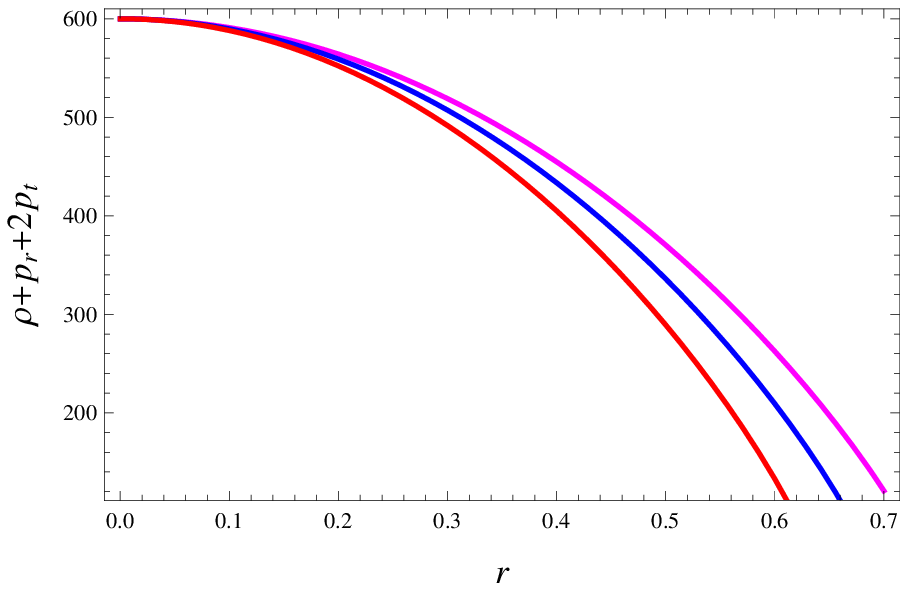,width=0.5\linewidth} \caption{Plots of energy
conditions versus radial coordinate for $p_{r}=\alpha\rho^{2}$,
$\beta=0.5$, $\alpha=0.0025$, $\sigma=2$, $\gamma=30$ (red),
$\gamma=35$ (blue) and $\gamma=40$ (magenta).}
\end{figure}

In the interior of polytropic star, the effect of anisotropic
pressure is examined by the anisotropic factor
$\Delta=\frac{2}{r}(p_{t}- p_{r})$. For $p_{t}> p_{r}$, $\Delta> 0$
indicating the outward direction of anisotropic pressure while for
$p_{t} < p_{r}$, $\Delta < 0$ which represents the inward direction
of anisotropic pressure. In order to obtain the expression of mass
function, the Misner-Sharp formula in GR is given by \cite{32}
\begin{equation}\label{23}
m=\frac{r}{2}\left(1+g^{ab}r_{,a}r_{,b}\right),
\end{equation}
where $m$ represents the gravitational mass within the sphere of
radius $r$. For static spherically symmetric spacetime, the
expression of mass function in GR is represented as
\begin{equation}\nonumber
m=\frac{r}{2}\left(1-e^{-\lambda}\right).
\end{equation}
Replacing this expression in Eq.(\ref{11}) yields
\begin{eqnarray}\nonumber
m'&=&\frac{r^{2}}{2(2\sigma R-\gamma\rho)}\left[\rho
+\frac{\gamma}{2}R\rho-\frac{\sigma}{2}R^{2}
+e^{-\lambda}\left\{\frac{\gamma}{8}\mu'^{2}\rho-\sigma R'
\left(\lambda'-\frac{4}{r}\right)\right.\right.\\\nonumber&+&\left.\left.\gamma
p_{r} \left(\frac{\lambda'}{r}-\frac{1}{r^{2}}\right)+2\sigma R''-
\frac{\gamma}{2} \rho'\left(\frac{\lambda'}{2}-\frac{5\mu'}{2}
-\frac{2}{r}\right)+\frac{\gamma}{2}\left(\rho''
-p_{r}''\right)\right.\right.
\\\nonumber&+&\left.\left.\frac{\gamma}{r}p_{t}'-\frac{\gamma}{2}p_{r}'
\left(\frac{4}{r}-\frac{\lambda'}{2}\right)\right\}\right].
\end{eqnarray}
This equation leads to the following form of mass function
\begin{eqnarray}\nonumber
m&=&\frac{1}{2}\int_{0}^{r}\frac{r^{2}}{2\sigma
R-\gamma\rho}\left[\rho +\frac{\gamma}{2}R\rho-\frac{\sigma}{2}R^{2}
+e^{-\lambda}\left\{\frac{\gamma}{8}\mu'^{2}\rho-\sigma R'
\left(\lambda'-\frac{4}{r}\right)\right.\right.\\\nonumber&+&\left.\left.\gamma
p_{r} \left(\frac{\lambda'}{r}-\frac{1}{r^{2}}\right)+2\sigma R''-
\frac{\gamma}{2} \rho'\left(\frac{\lambda'}{2}-\frac{5\mu'}{2}
-\frac{2}{r}\right)+\frac{\gamma}{2}\left(\rho''
-p_{r}''\right)\right.\right.
\\\label{24}&+&\left.\left.\frac{\gamma}{r}p_{t}'-\frac{\gamma}{2}p_{r}'
\left(\frac{4}{r}-\frac{\lambda'}{2}\right)\right\}\right]dr.
\end{eqnarray}
This is the Misner-Sharp formula for static spherically symmetric
spacetime corresponding to the $\sigma R^{2}+\gamma Q$ model of
$f(R,T,Q)$ gravity. For non-static spherically symmetric spacetime,
the Misner-Sharp formula in the same gravity is also discussed in
\cite{32a}. From Eq.(\ref{24}), we can check the behavior of masses
of anisotropic polytropes corresponding to two cases of polytropic
EoS with initial condition $m(0)=0$. The mass to radius ratio, also
known as compactness factor, is defined as
\begin{equation}\nonumber
u=\frac{m}{r}.
\end{equation}
The surface redshift also plays a dynamic role to understand the
physics of strong interaction between particles inside the star and
its EoS. The formula for $z_{s}$ relative to mass-function is
\begin{equation}\label{25}
z_{s}=\left(1-\frac{2m}{r}\right)^{-1/2}-1.
\end{equation}

The graphical interpretation of $\Delta$, mass-function, compactness
factor as well as redshift parameter is presented in Figure
\textbf{3}. The plot of anisotropic factor shows its positive
behavior yielding a repulsive force which permits the construction
of more massive distribution inside the polytropic star. Figure
\textbf{3} also illustrates that the value of mass-function remains
the same with increasing values of $\gamma$. In this case, the
maximum mass point of $0.19 M_{\odot}$ is obtained. The maximum
value of compactness factor is found to be less than $\frac{4}{9}$.
It is also observed that the bound of surface redshift is
$z_{s}\leq5.211$. This analysis shows that the values of $m$, $u$
and $z_{s}$ are in well agreement with the required limits.

\begin{figure}\center
\epsfig{file=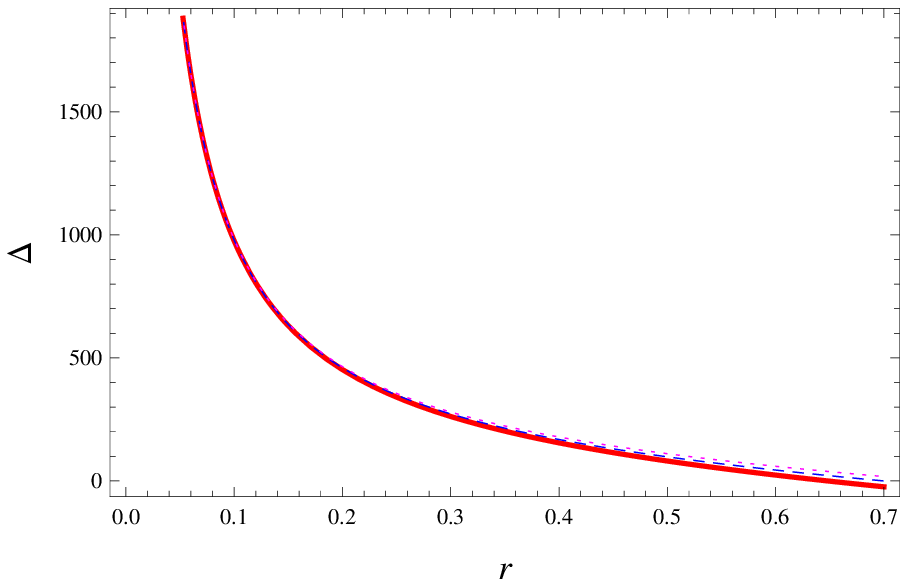,width=0.4\linewidth}\epsfig{file=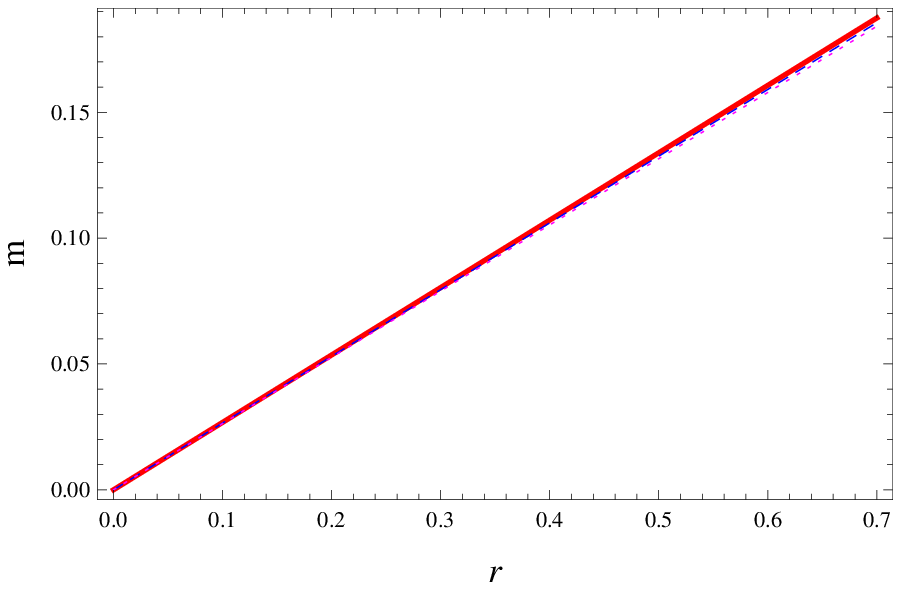,width=0.4\linewidth}
\epsfig{file=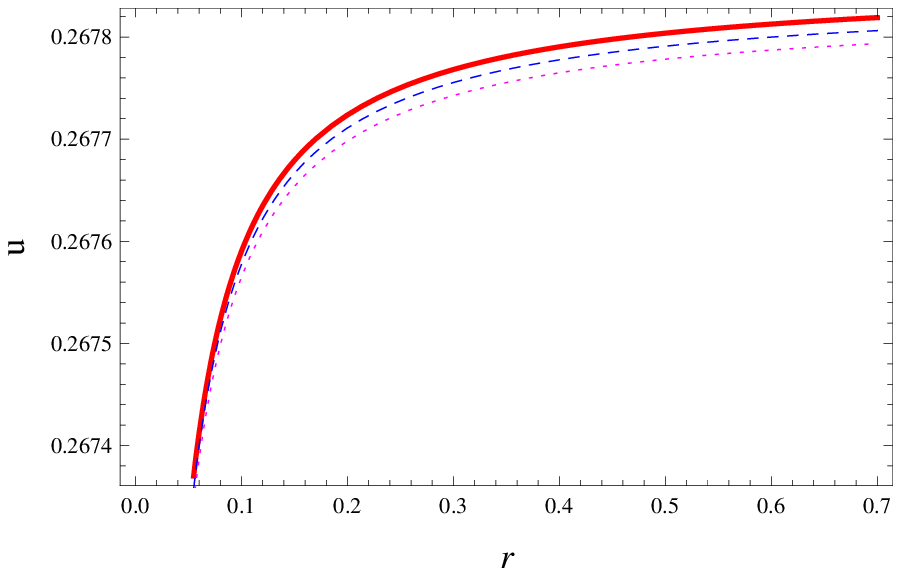,width=0.4\linewidth}\epsfig{file=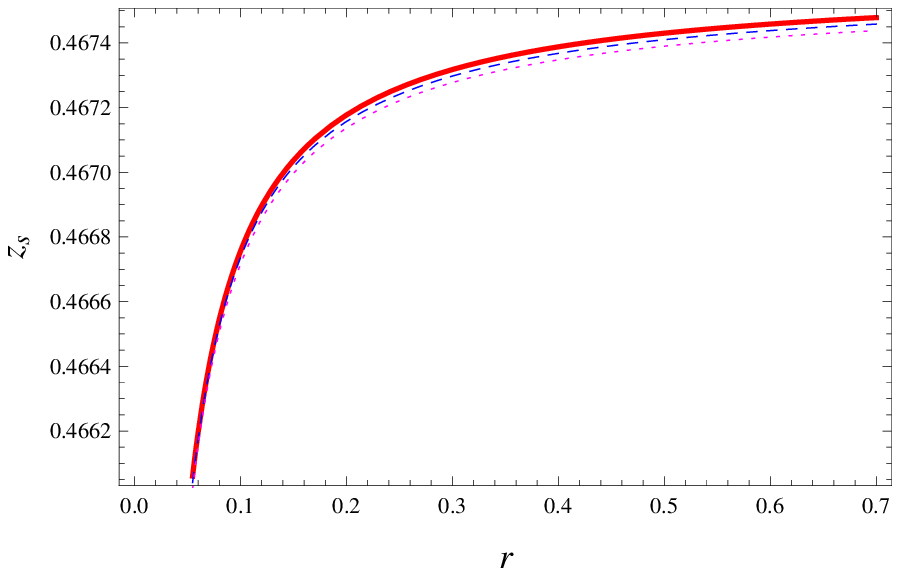,width=0.4\linewidth}
\caption{Plots of $\Delta$, $m$, $u$ and $z_{s}$ versus radial
coordinate for $p_{r}=\alpha\rho^{2}$, $\beta=0.5$, $\alpha=0.0025$,
$\sigma=2$, $\gamma=30$ (red), $\gamma=35$ (blue dashed) and
$\gamma=40$ (magenta dotted).}
\end{figure}

The stability of stellar structure has a great importance in
analyzing physically acceptable models. We investigate stability of
anisotropic polytropes through analyzing causality condition as well
as adiabatic index. According to causality condition, the squared
speed of sound defined by $v_{s}^{2}=dp/d\rho$ should lie within the
limit $[0,~1]$, i.e., $0\leq v_{s}^{2}\leq1$ everywhere in the
interior of stars for a physically stable stellar object. For
anisotropic fluid, we have $0\leq v^{2}_{sr}\leq 1$ and $0\leq
v^{2}_{st}\leq 1$, where $v_{sr}$ and $v_{st}$ stand for radial as
well as transverse components of sound speed, respectively. Herrera
\cite{33} introduced the concept of cracking using a different
approach to identify potentially stable/unstable structure of
compact objects. The potentially stable/unstable regions are
computed from the difference of sound speed in radial and transverse
directions.

Abreu et al. \cite{34} described the potentially stable and unstable
regions as
\begin{itemize}
\item $-1 \leq v^{2}_{st}-v^{2}_{sr} \leq 0$, for potentially stable
region,
\item $0 \leq v^{2}_{st}-v^{2}_{sr} \leq 1$, for potentially unstable region.
\end{itemize}
From $\mid v^{2}_{st}- v^{2}_{sr}\mid \leq 1$, we mean that there is
no cracking or overturning appears in the system. The adiabatic
index also plays a vital role to investigate the stability of
relativistic as well as non-relativistic stellar objects.
Chandrasekhar \cite{35} and other researchers \cite{36} studied the
dynamical stability against infinitesimal radial adiabatic
perturbation of the stellar system. It is found that the value of
adiabatic index should be greater than $\frac{4}{3}$ in the interior
of a dynamically stable stellar object. The expression of adiabatic
index for anisotropic fluid is defined by
\begin{equation}\label{26}
\Gamma_{r}=\frac{\rho+p_{r}}{p_{r}}\left(dp_{r}/d\rho\right),\quad
\Gamma_{t}=\frac{\rho+p_{t}}{p_{t}}\left(dp_{t}/d\rho\right).
\end{equation}
\begin{figure}\center
\epsfig{file=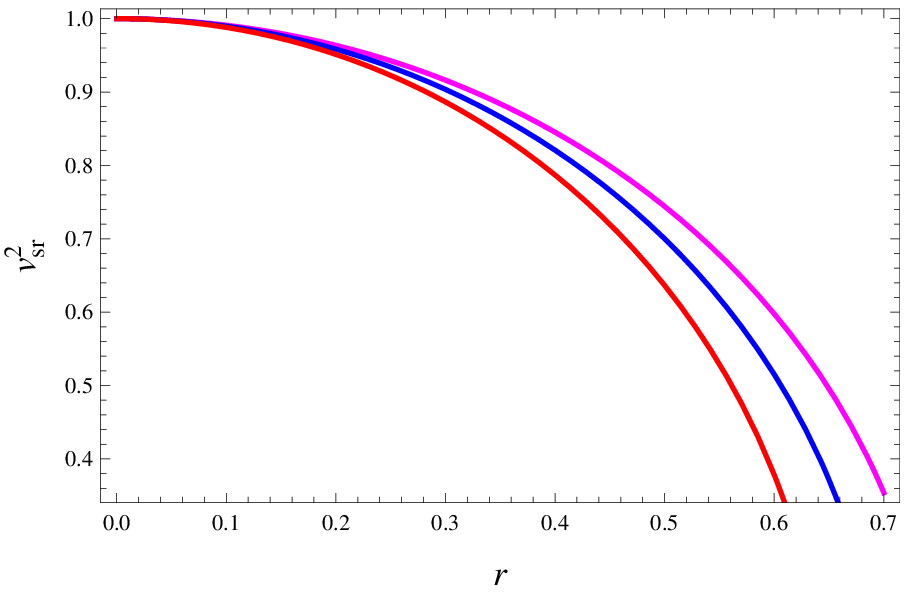,width=0.5\linewidth}\epsfig{file=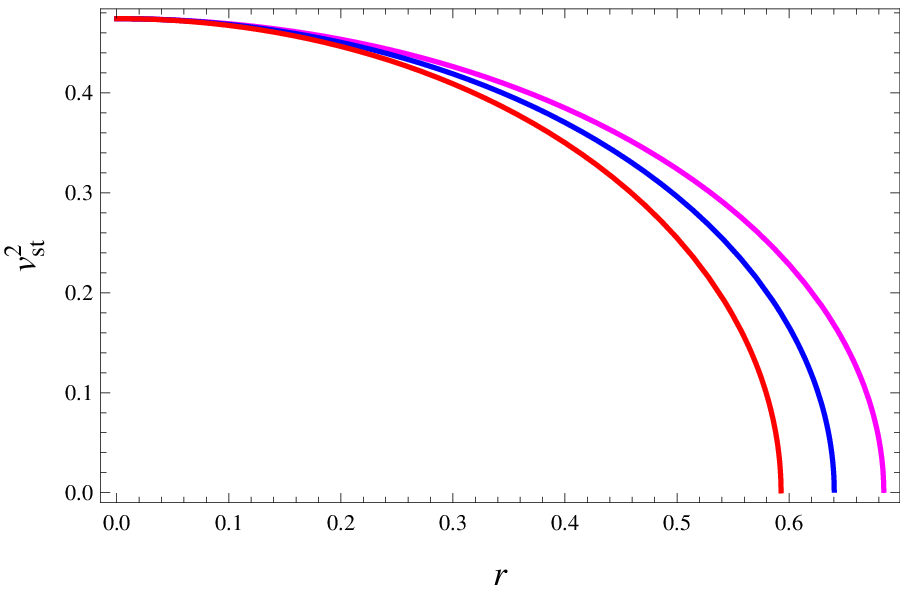,width=0.5\linewidth}
\epsfig{file=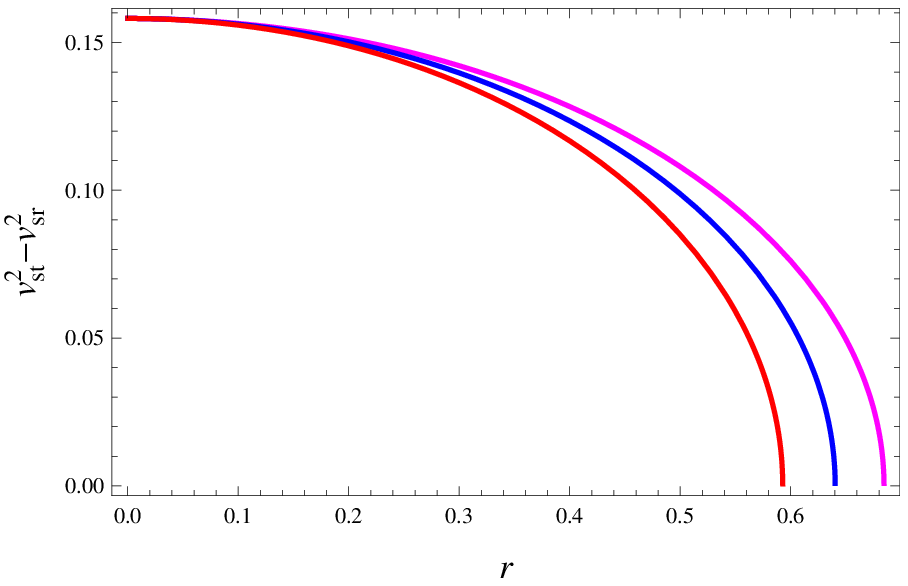,width=0.5\linewidth}\epsfig{file=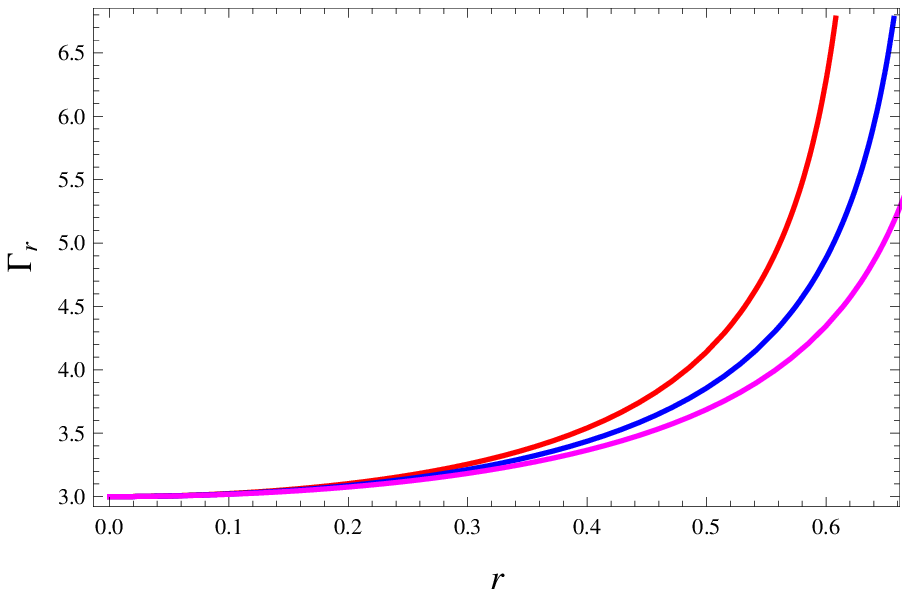,width=0.5\linewidth}
\epsfig{file=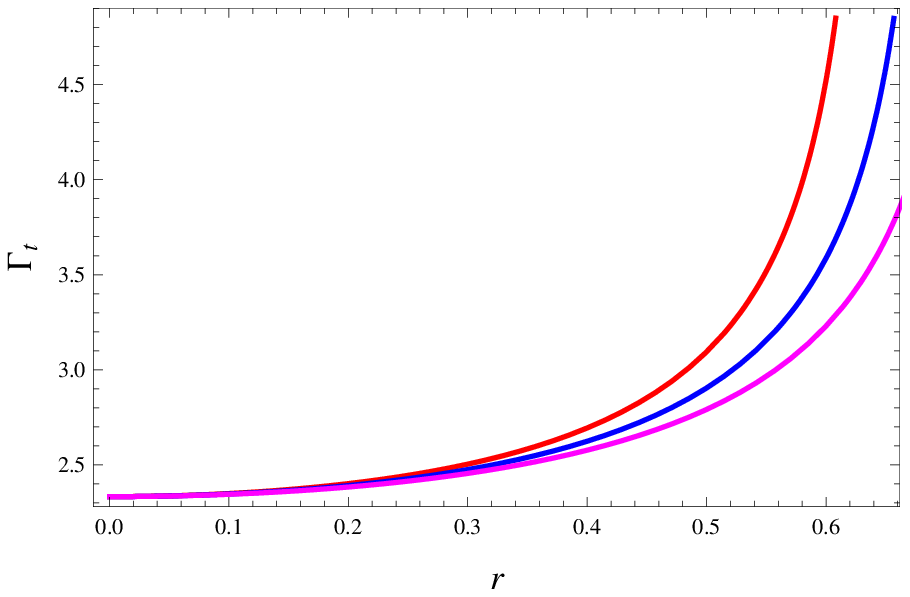,width=0.5\linewidth}\caption{Stability analysis
versus radial coordinate for $p_{r}=\alpha\rho^{2}$, $\beta=0.5$,
$\alpha=0.0025$, $\sigma=2$, $\gamma=30$ (red), $\gamma=35$ (blue)
and $\gamma=40$ (magenta).}
\end{figure}
Figure \textbf{4} represents the graphical behavior of squared speed
of sound and adiabatic index for polytropic EoS in this case. This
analysis indicates that the anisotropic polytropes are potentially
stable and no cracking is observed. It also shows that the system of
differential equations is dynamically stable for all values of
coupling parameter $\gamma$.

\subsection{Case II: $\nu=5/3$}

Here we explore the features of anisotropic polytropes considering
polytropic EoS $p_{r}=\alpha\rho^{\frac{5}{3}}$. The set of
Eqs.(\ref{12}) and (\ref{14})-(\ref{16}) corresponding to
Eq.(\ref{17}) turns out to be
\begin{eqnarray}\nonumber
p_{r}''&=&\frac{5\alpha^{\frac{3}{5}}
p_{r}^{\frac{2}{5}}}{3+5(1+\beta)\alpha^{\frac{3}{5}}
p_{r}^{\frac{2}{5}}}\left[p_{r}'\left\{\frac{3}{5\alpha^{\frac{3}{5}}
p_{r}^{\frac{2}{5}}}\left(\frac{\lambda'}{2}-\frac{5\mu'}{2}
-\frac{2}{r}\right)-2\beta\left(\frac{\mu'}{4}
-\frac{\lambda'}{4}\right.\right.\right.
\\\nonumber&+&\left.\left.\left.\frac{3}{r}\right)
-\frac{3\mu'}{2}-\frac{4}{r}+\frac{\lambda'}{2}
+\frac{6p_{r}'}{25\alpha^{\frac{3}{5}}p_{r}^{\frac{7}{5}}}\right\}
+p_{r}\left\{\frac{\lambda'}{r}-\frac{\mu'\lambda'}{2}+\frac{\mu'^{2}}{2}
+\frac{\beta\mu'}{r}\right.\right.\\\nonumber&+&\left.\left.\mu''-
\frac{2}{\alpha^{\frac{3}{5}}p_{r}^{\frac{2}{5}}}\left(\frac{\mu''}{2}
+\frac{\lambda'}{2r} +\frac{\mu'}{2r}-\frac{\mu'\lambda'}{4}
+\frac{e^{\lambda}}{r^{2}}-\frac{1}{r^{2}}+\frac{5\mu'^{2}}{8}\right)
-\frac{e^{\lambda}}{\gamma}(2\right.\right.\\\nonumber&+&\left.\left.R)\left(1
+\beta+\frac{1}{\alpha^{\frac{3}{5}}p_{r}^{\frac{2}{5}}}\right)\right\}
+\frac{4\sigma R}{\gamma}\left(\frac{\mu''}{2}+\frac{\lambda'}{2r}
+\frac{\mu'}{2r}-\frac{\mu'\lambda'}{4}+\frac{e^{\lambda}}{r^{2}}
-\frac{1}{r^{2}}\right.\right.\\\label{27}&+&\left.
\left.\frac{\mu'^{2}}{4}\right)-\frac{2\sigma
R'}{\gamma}\left(\frac{2}{r}-\mu'\right)\right],\\\nonumber
\lambda''&=&\lambda'\left(\frac{4\beta}{r}-\mu'
-\frac{3\lambda'}{2}-\frac{2p_{r}'}{p_{r}}+\frac{\mu'}{\alpha^{\frac{3}{5}}
p_{r}^{\frac{2}{5}}}\right)+e^{\lambda}\left(\frac{2}{r^{2}\alpha^{\frac{3}{5}}
p_{r}^{\frac{2}{5}}}-R-\frac{\sigma R^{2}}{\gamma
p_{r}}\right.\\\nonumber&-&\left.\frac{2}{\gamma}-\frac{4\sigma
R}{\gamma r^{2}p_{r}}\right)+\left(\frac{4\sigma R}{\gamma
rp_{r}}-\frac{2}{r\alpha^{\frac{3}{5}}
p_{r}^{\frac{2}{5}}}\right)\left(\mu'+\frac{1}{r}\right)
+\mu''-\frac{\mu'^{2}}{\alpha^{\frac{3}{5}}
p_{r}^{\frac{2}{5}}}-\frac{2\mu'}{r}\\\label{28}&+&
p_{r}'\left(\frac{3\mu'}{5\alpha^{\frac{3}{5}}p_{r}^{\frac{7}{5}}}
-\frac{2(1+\beta)}{rp_{r}}-\frac{1}{p_{r}}(\frac{2}{r}
+\frac{\mu'}{2})\right)+\frac{4\sigma R'}{\gamma p_{r}}(\frac{2}{r}
+\frac{\mu'}{2}),\\\nonumber \mu'''&=&\mu''\left(\frac{3\lambda'}{2}
-\frac{3\mu'}{2}-\frac{2}{r}-\frac{p_{r}'}{2p_{r}}
-\frac{\mu'}{2\alpha^{\frac{3}{5}} p_{r}^{\frac{2}{5}}}
-\frac{3p_{r}'}{10\alpha^{\frac{3}{5}}p_{r}^{\frac{7}{5}}}\right)
+\mu'\left\{\frac{\lambda''}{2}+\frac{2\lambda'}{r}\right.
\\\nonumber&-&\left.\frac{1}{\gamma}-\frac{\mu'^{2}}{4}
-\frac{\lambda'^{2}}{2} +\frac{3\mu'\lambda'}{4}
-R-\frac{1}{\gamma\alpha^{\frac{3}{5}} p_{r}^{\frac{2}{5}}}
+\frac{(1+\beta)p_{r}'}{rp_{r}}+\frac{2(1
+\beta)}{r^{2}}\right.\\\nonumber&-&\left.\frac{R}{\alpha^{\frac{3}{5}}
p_{r}^{\frac{2}{5}}}+\frac{p_{r}'}{4p_{r}}
(\lambda'-\mu')+\frac{1}{\alpha^{\frac{3}{5}}
p_{r}^{\frac{2}{5}}}\left(\frac{\mu'\lambda'}{4}
-\frac{\mu'}{r}-\frac{\mu'^{2}}{4}\right)-\frac{3p_{r}'}
{5\alpha^{\frac{3}{5}}p_{r}^{\frac{7}{5}}}\right.
\\\nonumber&\times&\left.\left(\frac{1}{r}
-\frac{\lambda'}{4}+\frac{\mu'}{4}\right)\right\}
-\frac{p_{r}'}{p_{r}}\left(\frac{2}{\gamma}+R
-\frac{\lambda'}{r}\right)+\frac{2\lambda''}{r}
+\frac{2\lambda'}{r^{2}}
+\frac{6Rp_{r}'}{5\alpha^{\frac{3}{5}}p_{r}^{\frac{7}{5}}}
\\\nonumber&+&R'\left(3+2\beta-\frac{1}
{\alpha^{\frac{3}{5}}
p_{r}^{\frac{2}{5}}}\right)-\frac{2\lambda'^{2}}{r}
+2(1+\beta)\left\{\frac{2}{r^{3}}-\frac{2e^{\lambda}}{r^{3}}
-\frac{\lambda'}{r^{2}}-\frac{p_{r}'}{p_{r}}\right.
\\\label{29}&\times&\left.\left(\frac{e^{\lambda}}{r^{2}}
-\frac{1}{r^{2}}+\frac{\lambda'}{2r}\right)\right\},\\\nonumber
R''&=&\frac{e^{\lambda}p_{r}}{6\sigma}\left\{3+2\beta-\frac{1}
{\alpha^{\frac{3}{5}}p_{r}^{\frac{2}{5}}}+\frac{\gamma
R}{2}\left(3+2\beta+\frac{1}{\alpha^{\frac{3}{5}}
p_{r}^{\frac{2}{5}}}\right)\right\}-R'\left(\frac{\mu'}{2}
+\frac{2}{r}\right.\\\nonumber&-&\left.\frac{\lambda'}{2}\right)
-\frac{\gamma}{12\sigma} p_{r}''\left(\frac{3}{5\alpha^{\frac{3}{5}}
p_{r}^{\frac{2}{5}}}-5-2\beta\right)+\frac{\gamma}{12\sigma}
p_{r}'\left(\frac{7\mu'}{2}+\frac{10}{r}-\frac{5\lambda'}{2}
\right.\\\nonumber&+&\left.\beta(\mu'-\lambda')
+\frac{6p_{r}'}{25\alpha^{\frac{3}{5}}p_{r}^{\frac{7}{5}}}+\frac{3}
{5\alpha^{\frac{3}{5}}p_{r}^{\frac{2}{5}}}\left(\frac{\lambda'}{2}
-\frac{2}{r}-\frac{\mu'}{2}\right)\right)
+\frac{\gamma}{6\sigma}p_{r}\left(\frac{\mu''}{2}
+\frac{\mu'}{r}\right.\\\label{30}&+&\left.\frac{\mu'^{2}}{4}
-\frac{\mu'\lambda'}{4}+\frac{\beta}{r}\left(\lambda'
-\mu'\right)+\frac{1}{\alpha^{\frac{3}{5}}
p_{r}^{\frac{2}{5}}}\left(\frac{\mu''}{2}-\frac{\mu'\lambda'}{4}
+\frac{\mu'}{r}+\frac{3\mu'^{2}}{4}\right)\right).
\end{eqnarray}

\begin{figure}\center
\epsfig{file=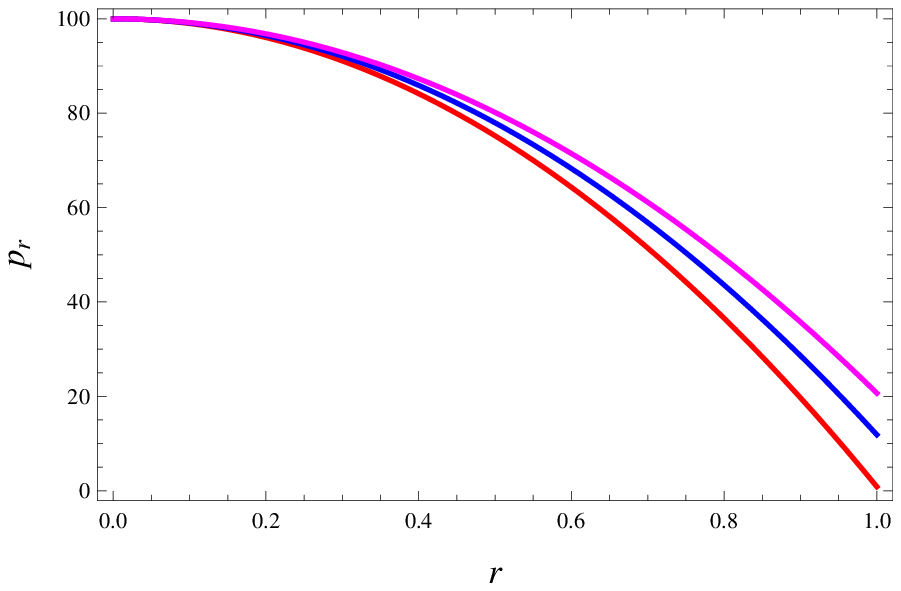,width=0.4\linewidth}\epsfig{file=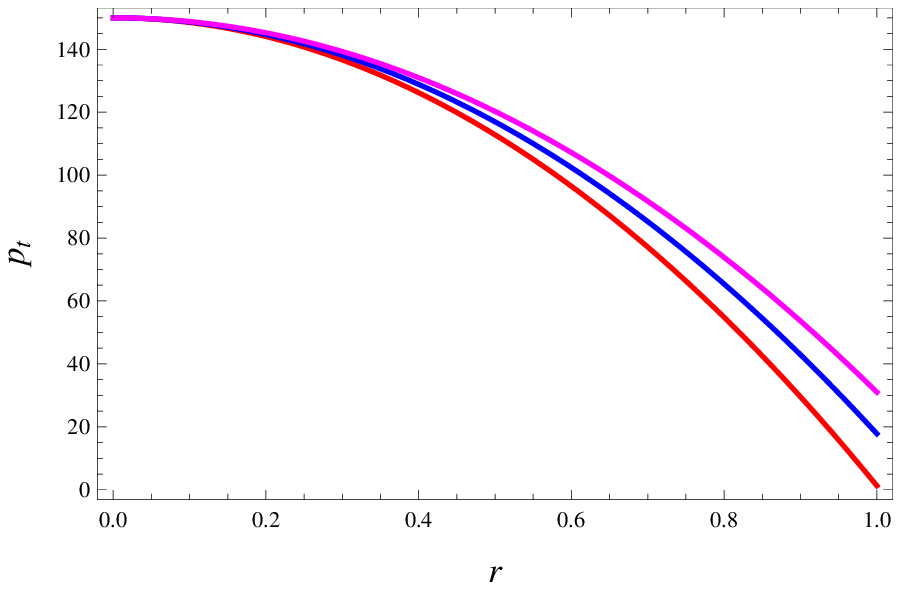,width=0.4\linewidth}
\epsfig{file=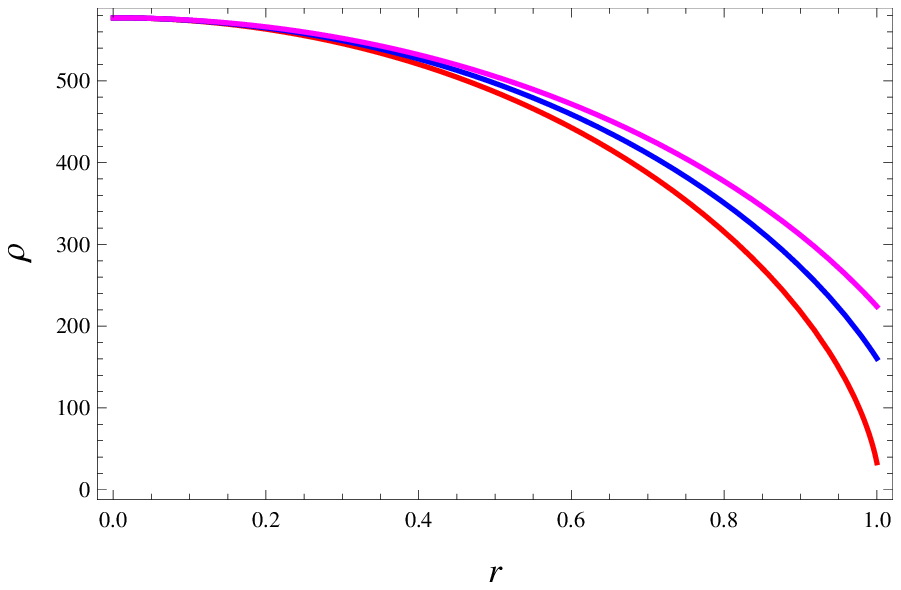,width=0.4\linewidth}\epsfig{file=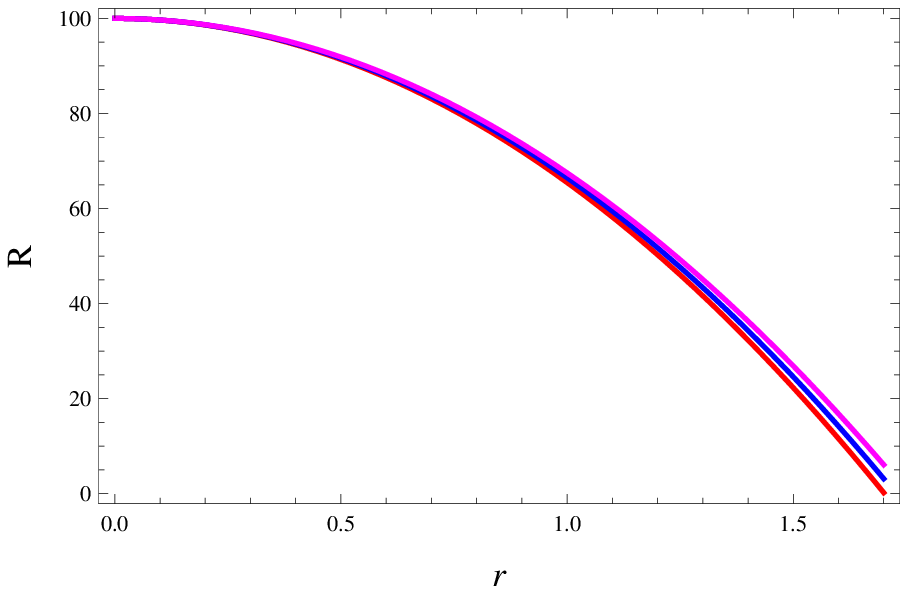,width=0.4\linewidth}
\caption{Variation of $p_{r}$, $p_{t}$, $\rho$ and $R$ versus $r$
for $p_{r}=\alpha\rho^{5/3}$, $\beta=0.5$, $\alpha=0.0025$,
$\sigma=2$, $\gamma=40$ (red), $\gamma=45$ (blue) and $\gamma=50$
(magenta).}
\end{figure}
\begin{figure}\center
\epsfig{file=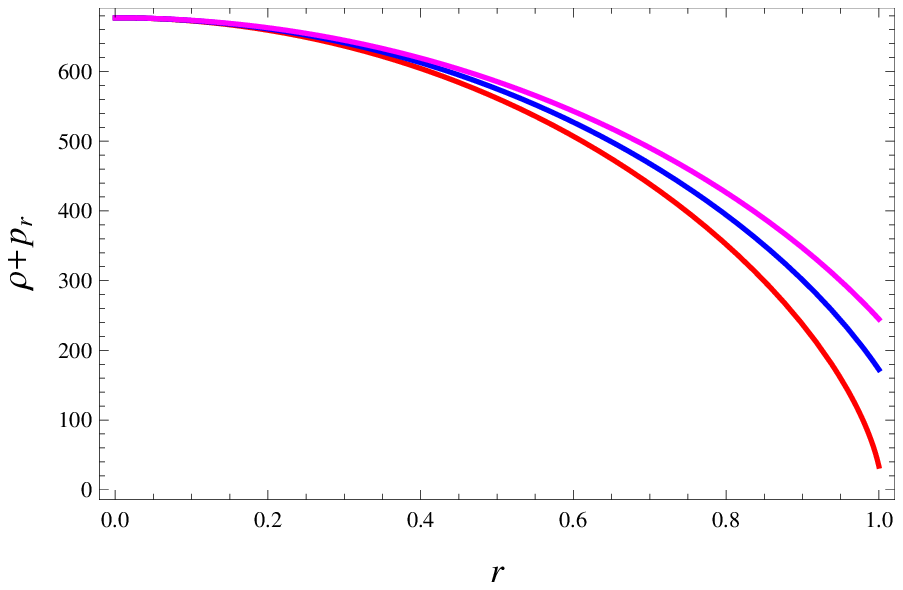,width=0.5\linewidth}\epsfig{file=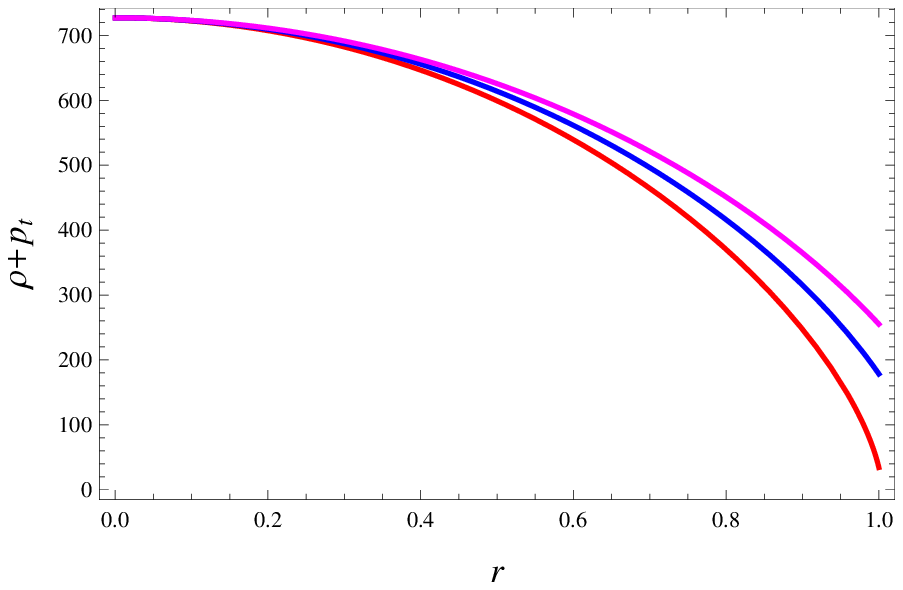,width=0.5\linewidth}
\epsfig{file=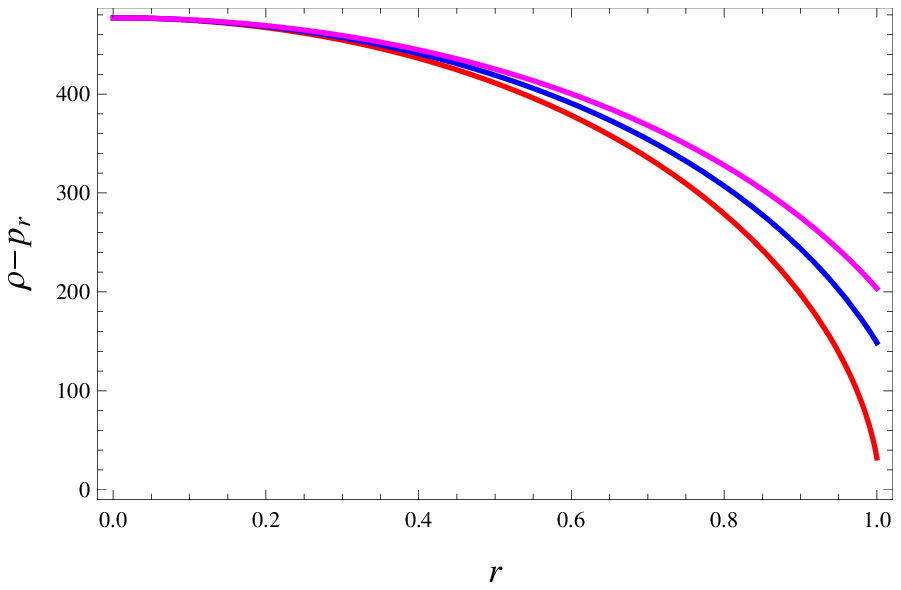,width=0.5\linewidth}\epsfig{file=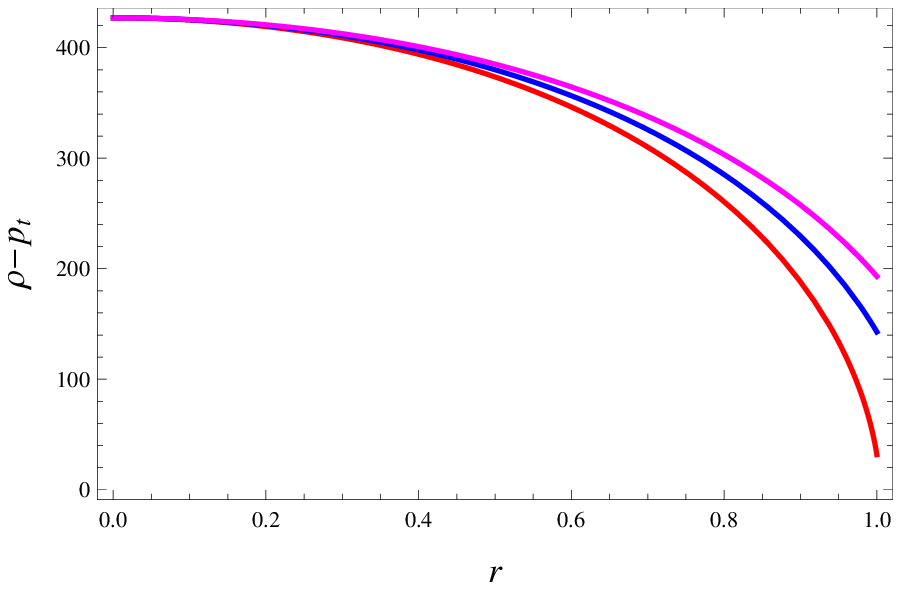,width=0.5\linewidth}
\epsfig{file=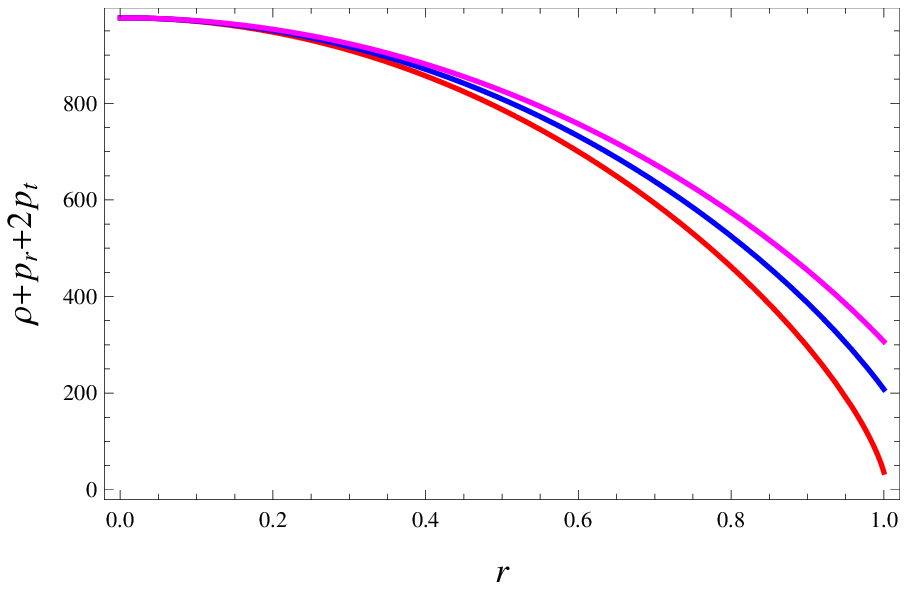,width=0.5\linewidth} \caption{Plots of energy
conditions versus radial coordinate for $p_{r}=\alpha\rho^{5/3}$,
$\beta=0.5$, $\alpha=0.0025$, $\sigma=2$, $\gamma=40$ (red),
$\gamma=45$ (blue) and $\gamma=50$ (magenta).}
\end{figure}
\begin{figure}\center
\epsfig{file=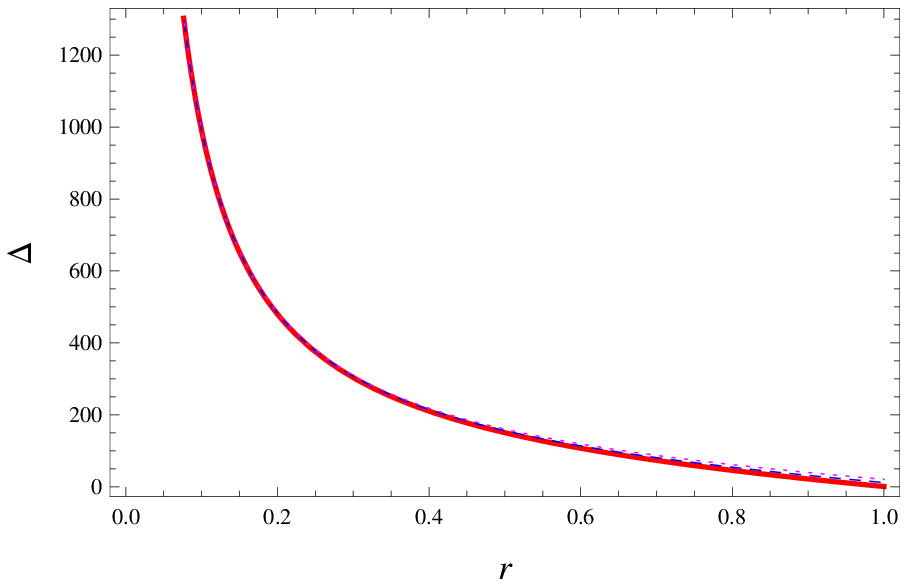,width=0.4\linewidth}\epsfig{file=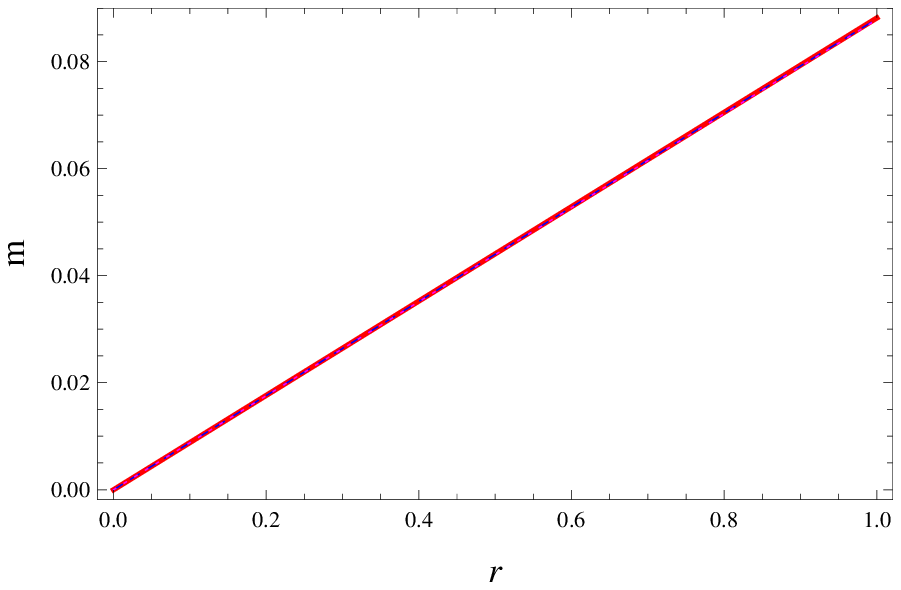,width=0.4\linewidth}
\epsfig{file=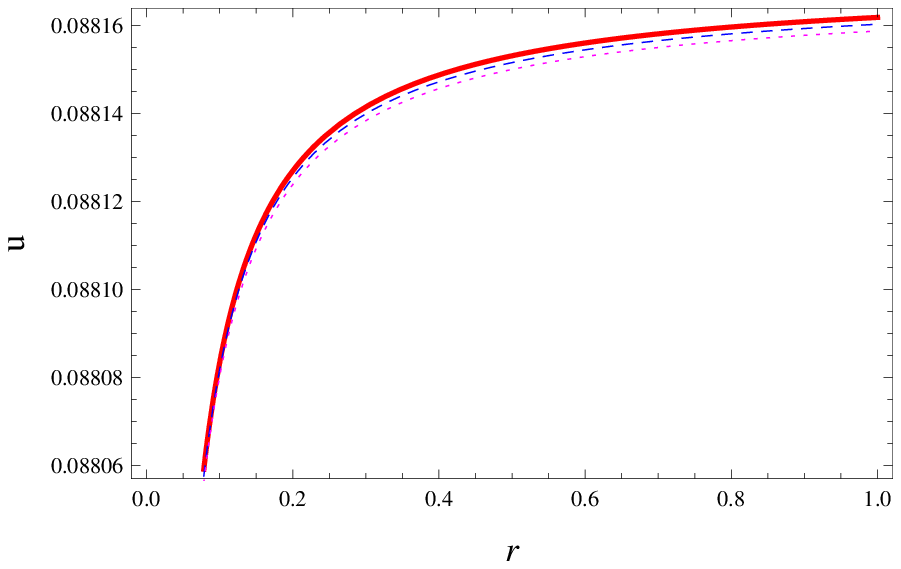,width=0.4\linewidth}\epsfig{file=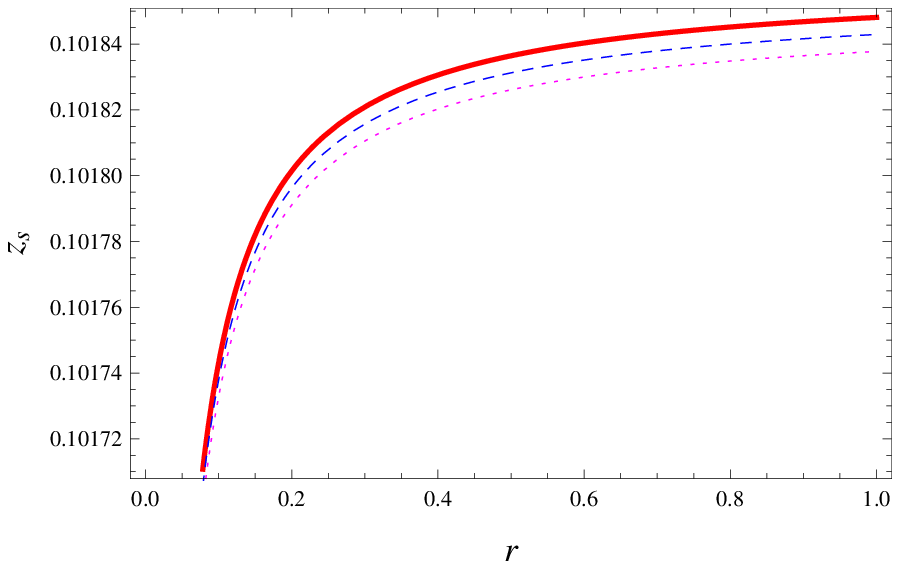,width=0.4\linewidth}
\caption{Plots of $\Delta$, $m$, $u$ and $z_{s}$ versus radial
coordinate for $p_{r}=\alpha\rho^{5/3}$, $\beta=0.5$,
$\alpha=0.0025$, $\sigma=2$, $\gamma=40$ (red), $\gamma=45$ (blue
dashed) and $\gamma=50$ (magenta dotted).}
\end{figure}
\begin{figure}\center
\epsfig{file=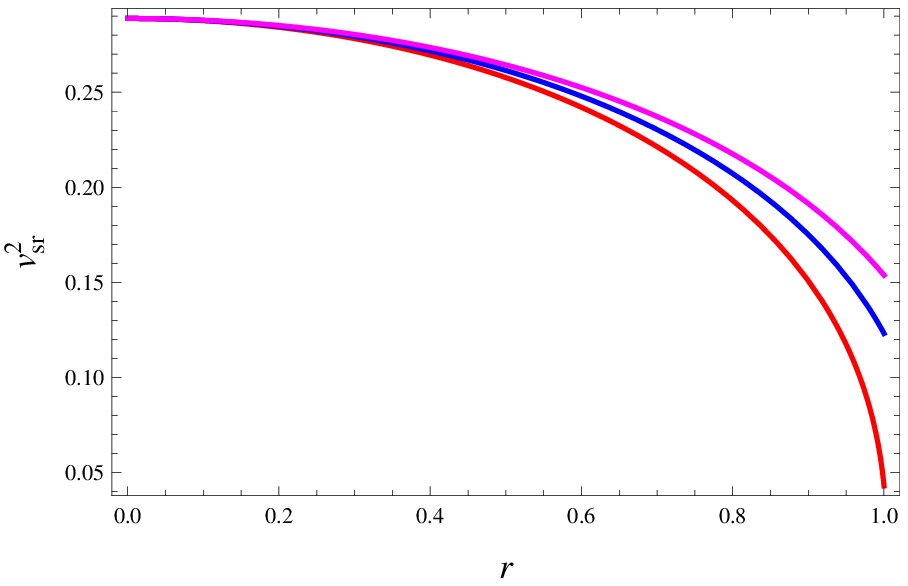,width=0.5\linewidth}\epsfig{file=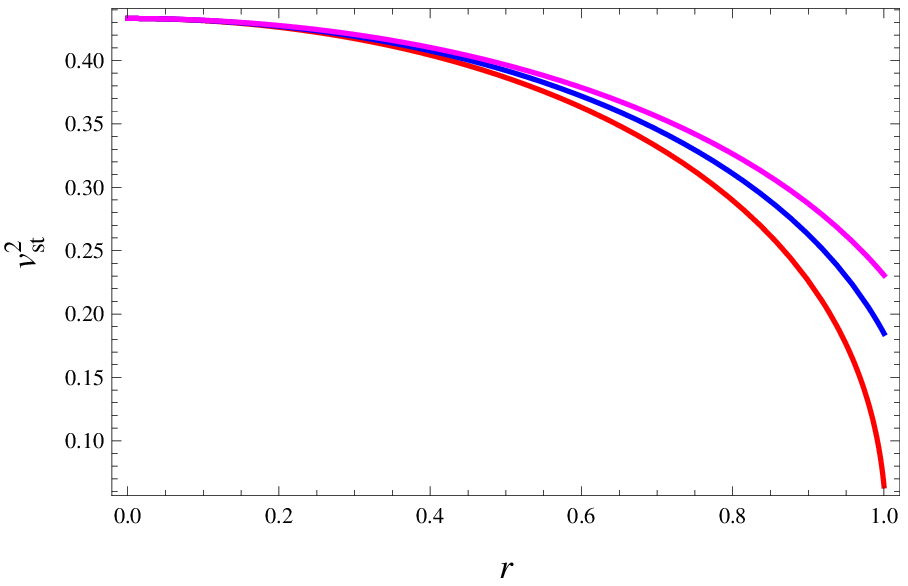,width=0.5\linewidth}
\epsfig{file=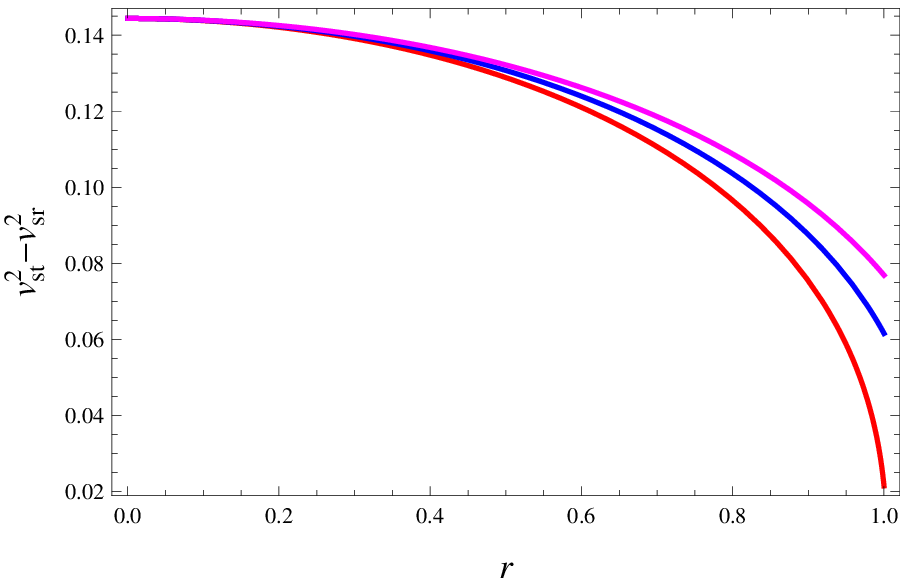,width=0.5\linewidth}\epsfig{file=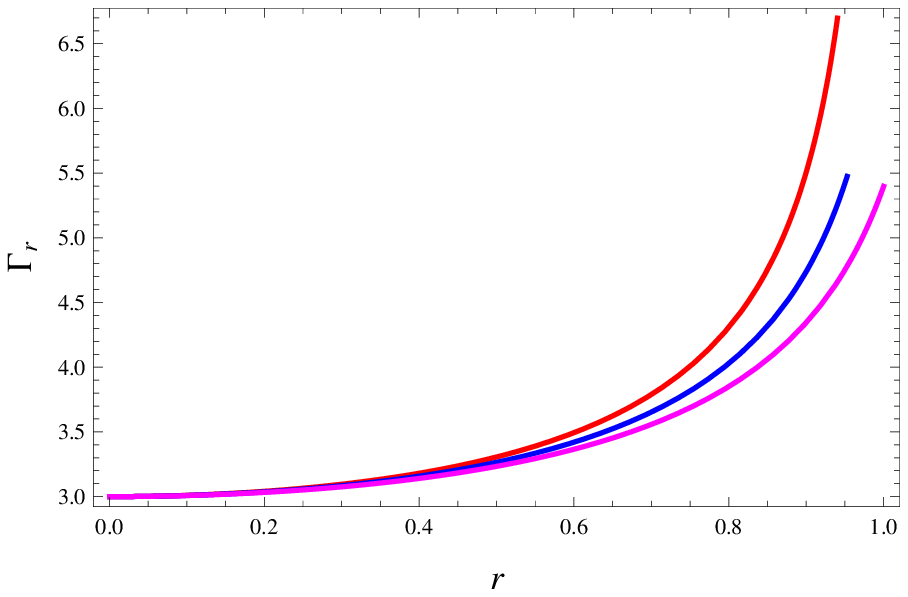,width=0.5\linewidth}
\epsfig{file=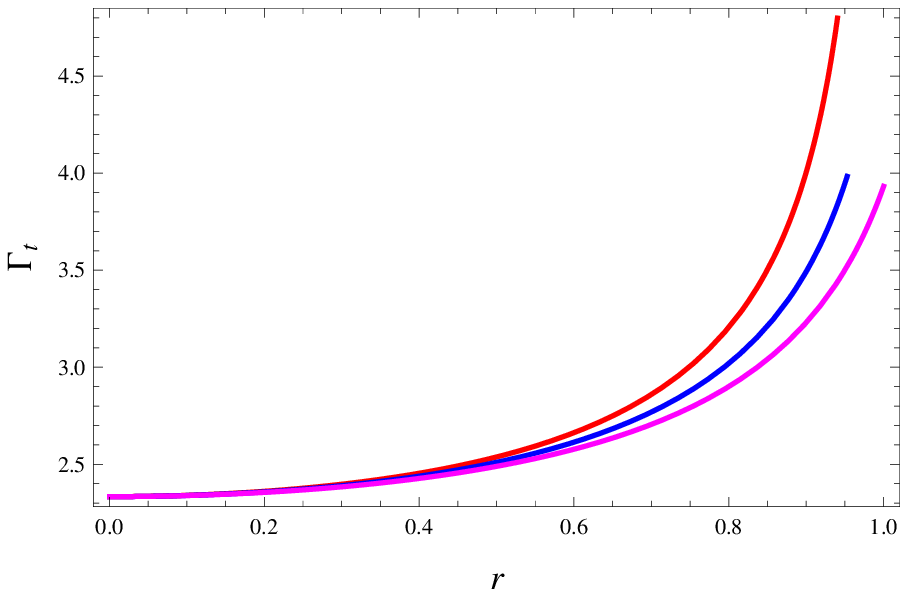,width=0.5\linewidth}\caption{Stability analysis
versus radial coordinate for $p_{r}=\alpha\rho^{5/3}$, $\beta=0.5$,
$\alpha=0.0025$, $\sigma=2$, $\gamma=40$ (red), $\gamma=45$ (blue)
and $\gamma=50$ (magenta).}
\end{figure}

We solve this system of equations numerically for the same initial
conditions, same values of $\alpha$, $\beta$ and $\sigma$ as in the
first case while the values of $\gamma$ are changed to obtain the
required behavior of density and pressure profiles. The graphs of
various physical quantities are plotted in Figures
\textbf{5}-\textbf{8}. The behavior shown in Figure \textbf{5}
indicates that the radial/tangential pressure, energy density and
Ricci scalar possess maximum values at the center that decrease
towards the boundary of polytropic star. However, in this case,
$p_{r}=0$ at $r=1 km$ suggesting that this is the radius of
anisotropic polytropes.

Figure \textbf{6} reveals that all energy conditions are also
satisfied corresponding to the second case of polytropic EoS. For
this case, the behavior of anisotropic factor is also positive,
maximum mass point of $0.1 M_{\odot}$ is found for all values of
coupling parameter $\gamma$, the value of compactness factor is
found to be less than $\frac{4}{9}$ and the value of gravitational
redshift also lies within the observational range as presented in
Figure \textbf{7}. The plots in Figure \textbf{8} represent that the
radial/tangential components of squared speed of sound lie within
the required range throughout the interior of stellar structure. The
squared tangential speed of sound is greater than squared radial
speed indicating no cracking occurs and $\mid v^{2}_{st}-
v^{2}_{sr}\mid \leq 1$ while adiabatic index is greater than
$\frac{4}{3}$ implies stable structure of polytropic star.

\section{Discussion and Conclusions}

The aim of this paper is to explore the basic physical
characteristics as well as stability of compact objects for two
cases of polytropic EoS in $f(R,T,Q)$ gravity. The analysis of
polytropic stars have been observed by constructing equations of
stellar structure, hydrostatic equilibrium equation and trace
equation under the effect of anisotropic pressure for particular
functional form $\sigma R^{2}+\gamma Q$ of this gravity. The
hydrostatic equilibrium equation also known as
Tolman-Oppenheimer-Volkoff equation, is an extension due to the
presence of extra terms coming from $\gamma Q$. The stellar
configurations of polytropic stars have been examined for different
values of the model parameter $\gamma$.

We have formulated a system of equations and developed the initial
conditions required for the numerical analysis. We have considered
two polytropic EoS $p_{r}=\alpha\rho^{2}$ and
$p_{r}=\alpha\rho^{5/3}$. The regularity conditions for energy
density, radial/tangential pressure are satisfied for both cases. It
is found that at the center, anisotropic polytropes exhibit maximum
pressure and density which decrease monotonically towards the
boundary of the star. The radii of approximately $0.7km$ and $1km$
are obtained for cases I and II, respectively. It is found that a
large value of coupling parameter $\gamma$ exhibits large radius of
the polytropic stars.

Our system of equations is also consistent with all energy
conditions corresponding to all chosen values of the model parameter
for both cases. The maximum mass points $0.19 M_{\odot}$ and $0.1
M_{\odot}$ with particular constants and initial conditions are
observed for the first and second polytropic EoS, respectively. The
maximum values of compactness factor are found 0.26782 for case I
and 0.08816 for case II. The maximum gravitational redshift for both
cases is $z_{s}\leq5.211$. We have found that radial/tangential
speed of sound for different values of $\gamma$ lie between $[0,~1]$
for both polytropic EoS which confirms the stable structure of
polytropic stars. The adiabatic index ($\gamma>\frac{4}{3}$) is
obtained for both cases which verifies the stability against an
infinitesimal radial adiabatic perturbation.

There are bounds on the masses of compact stars in GR. Chandrasekhar
\cite{37} suggested maximum mass of $1.4M_{\odot}$ for white dwarfs
to produce enough electron degeneracy pressure against collapse
whereas Tolman-Oppenheimer-Volkoff \cite{38} limit for the mass of
neutron stars is $3M_{\odot}$ to overcome gravity by neutron
degeneracy pressure. Also, in \cite{2} the anisotropic polytropes
with polytropic exponents 2 and 3/2 are discussed analytically and
the maximum mass point of $2.2976M_{\odot}$ for $R=7.07km$ is
observed. In \cite{4}, the charged polytropic stars are investigated
for different polytropic index and for $p_{r}=\alpha\rho^{2}$, the
maximum mass of $1.667M_{\odot}$ is obtained. In our scenario, we
have found the masses of $0.19 M_{\odot}$ and $0.1 M_{\odot}$ for
the first and second polytropic EoS, respectively which are less
than the masses observed in GR.

It is found that the masses, compactness factors and surface
redshift of the anisotropic polytropic stars lie within the
observational limits in $f(R,T,Q)$ gravity. We have also found that
the radii of anisotropic polytropes for $\sigma R^{2}+\gamma Q$
model are very small as compared to radii obtained for polytropic
EoS using perfect fluid distribution in $f(R,T)$ gravity \cite{23}.
However, these radii are comparable to those obtained for charged
polytropic stars corresponding to perfect fluid distribution
\cite{24}. We conclude that the effect of anisotropy as well as
non-minimal matter-curvature coupling provides a possibility of the
existence of such small compact objects.

\vspace{.25cm}

{\bf Acknowledgment}

\vspace{0.25cm}

We would like to thank the Higher Education Commission, Islamabad,
Pakistan for its financial support through the {\it Indigenous Ph.D.
5000 Fellowship Program Phase-II, Batch-III.}

\end{document}